\documentclass[fleqn,usenatbib]{mnras}

\usepackage{newtxtext,newtxmath}

\usepackage[T1]{fontenc}

\DeclareRobustCommand{\VAN}[3]{#2}
\let\VANthebibliography\thebibliography
\def\thebibliography{\DeclareRobustCommand{\VAN}[3]{##3}\VANthebibliography}

\usepackage{graphicx}	
\usepackage{amsmath}	
\usepackage{siunitx}
\usepackage{makecell}
\usepackage{array,multirow}
\usepackage{xurl}
\usepackage{tablefootnote}
\usepackage{float}
\usepackage{booktabs}

\title[]{A JWST investigation into the bar fraction at redshifts $1 \leq z \leq 3$}

\author[Z. A. Le Conte et al.]{Zoe A. Le Conte,$^{1}$\thanks{E-mail: zoe.a.le-conte@durham.ac.uk}
Dimitri A. Gadotti,$^{1}$
Leonardo Ferreira,$^{2}$
Christopher J. Conselice,$^{3}$
\newauthor
Camila de Sá-Freitas,$^{4}$
Taehyun Kim,$^{5}$
Justus Neumann,$^{6}$
Francesca Fragkoudi,$^{7}$
E. Athanassoula,$^{8}$
\newauthor
and
Nathan J. Adams$^{3}$
\\
$^{1}$Centre for Extragalactic Astronomy, Department of Physics, Durham University, South Road, Durham DH1 3LE, UK\\
$^{2}$Department of Physics \& Astronomy, University of Victoria, Finnerty Road, Victoria, British Columbia, V8P 1A1, Canada\\
$^{3}$Jodrell Bank Centre for Astrophysics, University of Manchester, Oxford Road, Manchester M13 9PL, UK\\
$^{4}$European Southern Observatory, Karl-Schwarzschild-Str. 2, D-85748 Garching bei Muenchen, Germany\\
$^{5}$Department of Astronomy and Atmospheric Sciences, Kyungpook National University, Daegu, 41566, Republic of Korea\\
$^{6}$Max-Planck-Institut f\"{u}r Astronomie, K\"{o}nigstuhl 17, D-69117 Heidelberg, Germany\\
$^{7}$Institute for Computational Cosmology, Department of Physics, Durham University, South Road, Durham DH1 3LE, UK\\
$^{8}$Aix Marseille Univ, CNRS, CNES, LAM, Marseille, France\\
}

\date{Accepted 28/03/2024. Received 05/03/2024; in original form 18/09/2023}

\pubyear{2023}

\begin{document}
\label{firstpage}
\pagerange{\pageref{firstpage}--\pageref{lastpage}}
\maketitle

\begin{abstract}
The presence of a stellar bar in a disc galaxy indicates that the galaxy hosts in its main part a dynamically settled disc and that bar-driven processes are taking place in shaping its evolution. Studying the cosmic evolution of the bar fraction in disc galaxies is therefore essential to understand galaxy evolution in general. Previous studies have found, using the Hubble Space Telescope (HST), that the bar fraction significantly declines from the local Universe to redshifts near one. Using the first four pointings from the James Webb Space Telescope (JWST) Cosmic Evolution Early Release Science Survey (CEERS) and the initial public observations for the Public Release Imaging for Extragalactic Research (PRIMER), we extend the studies of the bar fraction in disc galaxies to redshifts $1 \leq z \leq 3$, i.e., for the first time beyond redshift two. We only use galaxies that are also present in the Cosmic Assembly Near-IR Deep Extragalactic Legacy Survey (CANDELS) on the Extended Groth Strip (EGS) and Ultra Deep Survey (UDS) HST observations. An optimised sample of 368 close-to-face-on galaxies is visually classified to find the fraction of bars in disc galaxies in two redshift bins: $1 \leq z \leq 2$ and $2 < z \leq 3$. The bar fraction decreases from $\approx 17.8^{+ 5.1}_{- 4.8}$ per cent to $\approx 13.8^{+ 6.5}_{- 5.8}$ per cent (from the lower to the higher redshift bin), but is about twice the bar fraction found using bluer HST filters. Our results show that bar-driven evolution might commence at early cosmic times and that dynamically settled discs are already present at a lookback time of $\sim 11$ Gyrs.
\end{abstract}






\section{Introduction}
\label{Sec: intro}

Stellar bars are one of the most abundant features in local disc galaxies \citep[e.g.,][]{Eskridge_2000,Marinova_2007,Aguerri_2009,Buta_2015}, providing insight into the internal evolutionary processes taking place in these galaxies. Several investigations using optical surveys in the local Universe find strong stellar bars in about a third of disc galaxies \citep[e.g.,][]{Barazza_2009,Nair_2010b,Masters_2011}{}{}. This fraction increases to $60 - 80$ per cent if weaker bars are included \citep[e.g.,][]{Vaucouleurs_1991,Menéndez-Delmestre_2007,Sheth_2008,Erwin_2018}{}{}.

Barred stellar structures in disc galaxies are thought to form relatively quickly, over times of the order of a hundred million years, in massive disc galaxies which are dynamically cold and rotationally supported \citep[e.g.,][]{Hohl_1971,Kalnajs_1972,Ostriker_1973,Sellwood_1993}{}{}. Hence, the formation of stellar bars is an indicator of the evolutionary stage of a galaxy. The bar is a dense central region of evolved stellar populations on highly eccentric orbits \citep[e.g.,][]{Weinburg_1985,Contopoulos_1989,Athanassoula_1992,Kormendy_2004}{}{}. The non-axisymmetric nature of stellar bars is due to the very elongated form of the orbits that constitute the bar, which would be the $x_{1}$ orbital family or one of the higher multiplicity families, both parallel to the semi-major axis of the bar \citep{Contopoulos_1980,Wang_2022}, which have such properties. The orbital composition of the bar, coupled with the fact that the bar can be viewed from all possible angles, introduces a range of observed ellipticities. Therefore, the shape of the stellar bar in disc galaxies can appear shorter and more oval or longer and rectangular, thus influencing the bar strength measurement. The torque of the stellar bar redistributes the angular momentum within the galaxy \citep[e.g.,][]{Lynden_1972, Athanassoula_2003, Athanassoula_2005}{}{}. This makes bars a primary and efficient driver of internal evolution through the redistribution of baryonic and dark matter \citep[e.g.,][]{Menéndez-Delmestre_2007,Regan_2006,Matteo_2013,Fragkoudi_2018}{}{}. Bar-driven gas inflow considerably impacts central galactic star formation, most notably in the formation of stellar structures, such as the nuclear disc \citep[e.g.,][]{Sanders_1980,Knapen_1995,Allard_2006,Coelho_2011,Lorenzo_2012,Bittner_2020,Gadotti_2020}{}{}. The bar also undergoes buckling processes forming box/peanuts \citep[e.g.,][]{Combes_1981,Combes_1990,Ishizuki_1990,Kormendy_1982,Kormendy_2004,Carles_2016}{}{}. It is currently disputed as to whether the presence of a bar could influence the fueling mechanisms of the active galactic nucleus (AGN) although a consensus is emerging in that bars help building a fuel reservoir near the galactic centre \citep[e.g.,][]{Knapen_1995,Alonso_2013,Cisternas_2015,Alonso_2018,Silva-Lima_2022,Garland_2023}{}{}. \cite{Sheth_2005} confirm the result of \cite{Sakamoto_1999}, namely that the central kiloparsec of barred galaxies contains a higher degree of molecular gas concentrations, however in simulations \cite{Fragkoudi_2016} observe a consequential reduction in the gas inflow to the central kiloparsec due to the boxy/peanut bulge associated with the bar.

Multiple observational investigations into the abundance of stellar bars in disc galaxies up to $ z \simeq 1$ find a linear decrease in their frequency with increasing redshift. A constant bar fraction within the redshift range $0.25 < z < 1.0$ in the Galaxy Evolution from Morphologies and SEDs (GEMS) survey was found in \cite{Jogee_2004} where three independent techniques were used to identify spiral galaxies and ellipse fits were used to characterise barred galaxies. \cite{Abraham_1999} found a decline in the bar fraction within the redshift range $0.0 < z < 1.5$ from quantitatively estimated bar strengths of galaxies in the Hubble Deep Field-North and -South. \cite{Sheth_2003} identified barred galaxies by ellipse fitting techniques for galaxies within the redshift range $0.7 < z < 1.0$ in the Near-Infrared Camera and Multi-Object Spectrometer (NICMOS) Hubble Deep Field-North. Using the 2 deg$^{2}$ Cosmic Evolution Survey (COSMOS), \cite{Sheth_2008} found a decrease in the bar fraction using cross-checked visual and ellipse fitting bar identification techniques within the redshift range $0.20 < z < 0.84$. A decrease by a factor of two from $z \sim 0.4$ to $z \sim 1.0$ in the COSMOS bar fraction was found in \cite{Melvin_2014} using visual classifications. It has then been inferred from these studies that bar features cease to exist at greater lookback times, implying that bar-driven evolutionary processes do not commence until $\sim 6$ Gyr after the Big Bang. These studies require high-resolution and sensitive imaging across a large sky area, which the Hubble Space Telescope (HST) has achieved. At $z \simeq 1.5$, \cite{Simmons_2014} discover prominent bars in massive disc galaxies and suggest that at $\sim z > 1$, the bar fraction is sustained at $\sim$ 10 per cent. Two observational studies of the evolution of the bar fraction with redshift find no sign of a sharp decline at $z > 0.7$: \cite{Elmegreen_2004} find a near constant bar fraction of $0.23 \pm 0.03$ at redshifts from $z = 0$ up to $z \simeq 1.1$ for a sample of 186 disc galaxies; \cite{Jogee_2004} find the optical bar fraction of $\sim 0.3 \pm 0.06$ to remain at redshifts $0.2 < z < 1.0$.

In cosmological simulations, \cite{Kraljic_2012} found a depletion in the number of bars in present-day spiral progenitors at $0 < z < 2$, implying a violent phase of galaxy evolution where discs are dynamically hot, and there are excessive merger events. However, \cite{Athanassoula_2016} follow the merging of two disc galaxies and found that the merger remnant starts forming a bar before the disc is fully developed. \cite{Rosas_Guevara_2022} use TNG50 simulations \citep[][]{Nelson_2019}{}{} to trace the bar fraction evolution with redshift and show the bar fraction to increase to $\sim 50$ per cent at $z \simeq 1$ and only significantly decrease at $\sim z > 2$. Even at $z \simeq 6$, the simulated bar fraction, at a minimum, reaches $\sim$ 25 per cent. The bar fraction found in the Auriga cosmological zoom-in simulations from \cite{Fragkoudi_2020} are in good agreement with observational studies, where for redshifts $0 \leq z \leq 1.5$ the bar fraction decreases from $\sim 70$ per cent to $\sim 20$ per cent.

Various bar identification techniques can be applied to images, including classifications by eye \citep[e.g.,][]{Athanassoula_1990,Cheung_2013,Simmons_2014,Buta_2015}{}{}. Stellar bar characterisation and analysis identified structural features by eye from the colour composite images of galaxies, where participants vote a galaxy as \textit{barred, candidate bar} or \textit{unbarred} \citep[e.g.,][]{Vaucouleurs_1991,Eskridge_2000,Nair_2010b,Buta_2015}{}{}. Characteristic signatures in the radial profiles of barred galaxies can be seen, which can be used to aid or replace visual classification methods. Position angle (PA) and ellipticity ($e$) measurements are obtained from isophotal ellipse fits (see \S~\ref{Sec: criterion} for an explanation), in which the parameter radial profiles are used to identify a bar feature. The criteria for bar identification differ between studies but generally agree that within the bar-dominated region, the PA remains constant, and $e$ gradually rises. The end of the bar can be defined by taking the radius of either the peak ellipticity, or the one with the minimum ellipticity succeeding the peak ellipticity, or where a significant change in PA occurs, or a combination of these three metrics \citep[e.g.,][]{Wozniak_1995,Buta_1998,Elmegreen_2004,Jogee_2004,Marinova_2007,Guo_2023}{}{}. In a volume-limited $z \leq 0.01$ SDSS/DR7 sample with galaxies $M_{r} \leq$ -15.2, \cite{Lee_2019} found the bar fractions for three different identification techniques: 63\% by visual inspection; 48\% by ellipse fitting; 36\% by Fourier analysis. Additionally, in their study, they concluded that ellipse fitting techniques could miss $\sim 15\%$ of visually classified bars due to large bulges in early-type spirals. Using a deep convolutional neural network, \cite{Abraham_2018} identified bars in SDSS with good accuracy. Surveys are now on remarkably large scales, so automated techniques such as machine learning \citep[e.g.,][]{Cheng_2021}{}{} will become vital for morphological classifications.

The James Webb Space Telescope (JWST) has provided the opportunity to expand the investigation of the bar fraction to higher redshifts. Imaging from the Near Infrared Camera (NIRCam) probes the rest-frame near-infrared (NIR) emission of galaxies at redshifts up to 3 and probes the rest-frame optical at redshifts up to 7; NIR emission traces the older stellar populations which dominate bar features and are also less affected by dust extinction and recent star formation. \citep[e.g.,][]{Frogel_1996,Schneider_2006}{}{}. In fact, the NIR bar fraction at $z \simeq 0$ is higher than the optical bar fraction \citep[e.g.,][]{Marinova_2007}{}{}, and \cite{Buta_2015} argues that this is due to stellar structural features being more perceptible. Thus, weaker bars in the optical become stronger in the NIR, so a higher bar fraction is observed. In addition, the primary mirror on JWST is over 2.5 times the diameter size of the HST primary mirror, meaning that the sensitivity of JWST is significantly better. The improved sensitivity, along with the longer rest-frame wavelengths probed by JWST, means elongated bar structures become more discernible than in their counterpart HST images \citep[e.g.,][]{Huertas-Company_2023}{}{}. For this reason, we can now study the bar-driven evolution of galaxies with the JWST by searching for the epoch when stellar barred structures form in disc galaxies. 

By conducting visual classifications on NIRCam F200W filter images of the SMACS0723 cluster at $z = 0.39$, \citet{Mendez_2023} find that the bar fraction distribution is strongly dependent on stellar mass. A previous study of stellar bars at $\sim z > 1$ using the Cosmic Evolution Early Release Science Survey (CEERS) was conducted by \cite{Guo_2023} who identified six strongly barred galaxies at $z \sim 1 - 3$, with the highest redshift galaxy at $z \sim 2.3$. In this study, we use the initial four NIRCam JWST observations from CEERS to find the evolution of the bar fraction at redshifts between $z = 1 - 3$. To this aim, we visually classify a mass-complete sample of these high-resolution rest-frame NIR images for barred features in disc galaxies. 

This paper is outlined as follows: in \S~\ref{Sec: selection}, we explain the NIRCam image reduction pipeline and our sample selection. Stellar bar identification techniques and our methodology for visual classifications are discussed in \S~\ref{Sec: identification}. In \S~\ref{Sec: fraction}, we present the bar fraction for two redshift bins, $z = 1 - 2$ and $z = 2 - 3$, \S~\ref{Sec: discussion} discusses the implications of our findings on when bar-driven evolution commences and, summarise our results in \S~\ref{Sec: conclusion}. Throughout this study, we assume the latest Planck flat $\Lambda$CDM cosmology with H$_{0}$ = 67.36, $\Omega_{m}$ = 0.3153, and $\Omega_{\Lambda}$ = 0.6847 \citep{Planck_2020}.



\section{The parent sample}
\label{Sec: selection}
To define our sample, we use the initial four public NIRCam JWST observations from the Cosmic Evolution Early Release Science Survey (CEERS; PI: Filkelstein, ID=1345, \citealt{Finkelstein_2023}, CEERS1, CEERS2, CEERS3 and CEERS6) taken in June 2022 that overlap with the Cosmic Assembly Near-IR Deep Extragalactic Legacy Survey (CANDELS; \citealt{Grogin_2011, Koekemoer_2011}) on the Extended Groth Strip field (EGS), as well as the initial public observations for the Public Release Imaging for Extragalactic Research (PRIMER; PI: Dunlop, ID=1837, \citealt{DUNLOP2021PRIMER}), that overlap with the CANDELS Ultra Deep Survey (UDS) Field observations. Together, the data covers $\sim \ 30$ arcmin$^2$ of an area with CANDELS HST overlap. 

\subsection{Data Reduction Pipeline}
\label{sec:reduction}

We reprocess all of the uncalibrated lower-level JWST data products following a modified version of the JWST official pipeline. This is similar to the process used in \citet{Adams_2023} and exactly the same reductions as used in \citet{Ferreira_2022}, which can be summarised as follows: (1) We use version 1.6.2 of the pipeline with the Calibration Reference Data System (CRDS) version 0942 which was the most up-to-date version at the time these data products were generated. Use of CRDS 0942 is essential for zero point issues as discussed in \citet{Adams_2023}. (2) We apply the 1/f noise correction derived by Chris Willott on the resulting level 2 data of the JWST pipeline.\footnote{\url{https://github.com/chriswillott/jwst}} (3) We extract the sky subtraction step from stage 3 of the pipeline and run it independently on each NIRCam frame, allowing for quicker assessment of the background subtraction performance and fine-tuning. (4) We align calibrated imaging for each individual exposure to GAIA using \texttt{tweakreg}, part of the DrizzlePac python package.\footnote{\url{https://github.com/spacetelescope/drizzlepac}} (5) We pixel-match the final mosaics with the use of \texttt{astropy reproject}. The final resolution of the drizzled images is 0.03 arcseconds/pixel. 

Furthermore, an additional step was added for the PRIMER reductions in step (2) above due to the presence of a significant stripping pattern artefact at a 45-degree angle in the NIRCam footprint, resembling the diffraction pattern of a bright star outside the field of view of the camera. This issue was removed with an adaptation of the 1/f noise algorithm, first rotating the observations to 45 degrees to align the pattern with one of the axes, followed by a background subtraction for each row based on the background mean of that row. Finally, the adjusted file is rotated back to its original orientation. This drastically reduces the artefact in the final products, although some are still visible in colour composites due to the non-uniform nature of the artefact across different NIRCam filters. Galaxy stamps that present these residual artefacts are flagged during subsequent classification as described in \S~\ref{Sec: visual}.

Each one of the four June CEERS observations was processed into individual mosaics, while the PRIMER UDS observations were stacked in a single mosaic due to the large overlapping area.

\subsection{Sample Selection}
\label{sec:sampleselection}

As a way to produce a selection with robust photometric redshifts and stellar masses, we use the CANDELS-based catalogues produced by \citet{Duncan_2019} that include observations from HST, Spitzer, and ground-based facilities. These redshifts are robustly calibrated from spectroscopic redshifts, with an average outlier fraction of $\frac{|\Delta z|}{1+z_{spec}} \sim 5\%$ (see \citealt{Duncan_2019} for details).

From these catalogues, we select all sources that lie within the footprint of the CEERS and PRIMER observations outlined previously. All sources with photometric redshifts and stellar masses that are present in both CANDELS and the new JWST observations are selected. Additionally, no magnitude or signal-to-noise cut is done to mitigate any selection bias due to different sensitivities between HST and JWST, which prevents JWST bright galaxies from being excluded if they are faint in HST bands. Then, all overlapping sources between $1 \leq z \leq 3$ are selected, resulting in a parent sample of $5218$ galaxies present within the combined area of CEERS+PRIMER and $5445$ galaxies in the area of CANDELS EGS and UDS fields, including $3559$ galaxies with visual Hubble type classifications from \citet{Ferreira_2022} at $z > 1.5$. We note that some galaxies fall in the gaps between the NIRCam detectors, and therefore, while they are included in the HST sample, they cannot be analysed with JWST data.

For each of the CEERS+PRIMER $5218$ galaxies in the sample, we produce $30 \ \rm mas$ 128x128 pixel cutouts for the JWST filters, namely F356W and F444W. Concerning the 5445 CANDELS galaxies observed with the HST Wide Field Camera 3 filters, namely F160W, we produce $60 \ \rm mas$ 64x64 pixel cutouts covering a consistent angular field of view, enabling us to probe the same galaxies in a relatively similar wavelength regime between the two instruments. 
In this study, we select the F444W JWST filter and compare these galaxies to their HST WFC3 filter F160W. 



\section{Bar identification}
\label{Sec: identification}

The random orientation of galaxies challenges observational attempts of bar measurements. Stellar bars are distinguishable in near-face-on galaxies and become less well defined in high-inclination galaxies. This study aims to determine the fraction of disc galaxies that harbour a bar in an optimised sample of F444W NIRCam and F160W WFC3 images. For our bar identification process, we use visual inspection of galaxy images as well as radial profiles of position angles and ellipticity. 


\subsection{Sample optimisation}
\label{Sec: criterion}

\begin{figure*}
    \centering
    \includegraphics[width=\textwidth]{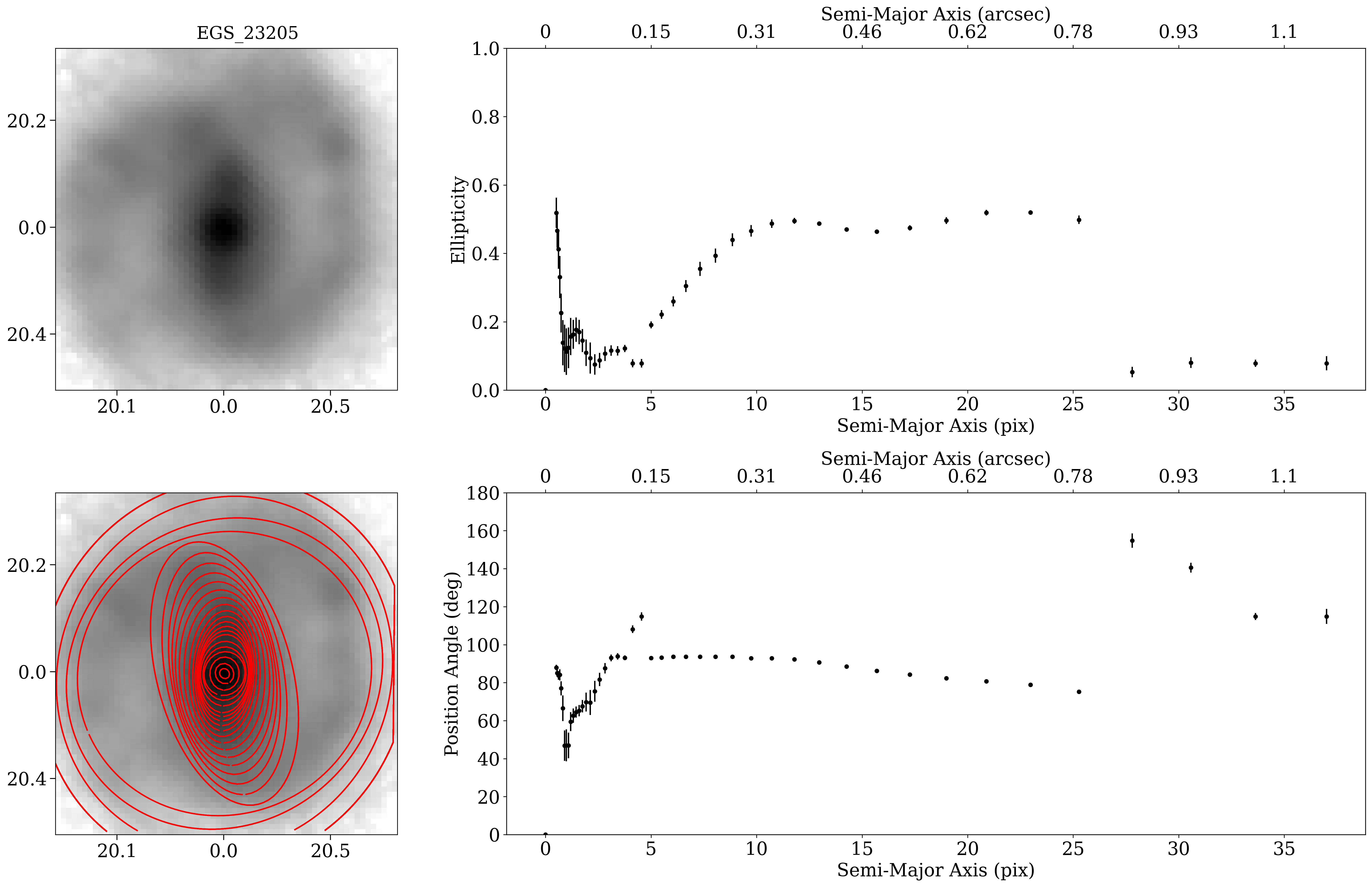}
    \caption{Elliptical isophotal fits using the module \texttt{photutils.isophote} from Python's astropy package \citep{Bradley_2022} to logarithmic F444W NIRCam images of the galaxy EGS\_23205 at redshift $z \sim 2.12$. The left-hand side shows the F444W image annotated with the pixel coordinates (top) and superposed elliptical isophotal fits (bottom). The right-hand side shows radial profiles of the ellipticity ($e$) (top) and position angle (PA) in degrees (bottom) as derived from the ellipse fits.}
    \label{Fig: isophotes}
\end{figure*}

Considering the challenges involved in the identification of bars, as noted above, we choose to remove highly inclined and overly faint or poorly resolved galaxies from the sample through an automated process. This optimisation process is intended to remove only galaxies that would be visually classified as ambiguous or unclassifiable. To do so, we fit ellipses to the isophotal contours of all galaxies in the parent sample to extract radial profiles of $e$ and PA \citep[see, e.g.,][]{Gadotti_2007,Barraza_2008,Barazza_2009,Aguerri_2009}{}{}. Figure \ref{Fig: isophotes} shows in the left panel the ellipse fits superposed on the F444W NIRCam image of the galaxy EGS\_23205, and in the right panel, radial profiles of the $e$ and PA of the fitted ellipses. EGS\_23205 is an example of a barred galaxy in this study and is observed relatively face-on.

Before visually classifying galaxies as barred or unbarred, we apply a three-step procedure to obtain our final, optimised galaxy sample containing galaxies in which a bar can be identified robustly: \textbf{(1)} ellipse-fitting to NIRCam images without fixing the centre; \textbf{(2)} second ellipse-fitting with fixed centres; \textbf{(3)} removal of highly inclined galaxies. In the following, we give a detailed explanation of these three steps:

\begin{description}
    \item[\textbf{Phase 1}] Elliptical isophotes are fitted to F444W NIRCam images of the JWST galaxy sample and analysed using \texttt{photutils.isophote} from Python's astropy package \citep{Bradley_2022}. This package uses an iterative method to measure the isophotes \citep{Jedrzejewski_1987}. The objective at this optimisation phase is to eliminate the overly disturbed or extremely poorly resolved or low surface brightness galaxies that are visually unclassifiable. In general, the algorithm fails to produce any result when applied to these objects. Therefore, at this stage, all galaxies for which fits can be obtained successfully are kept, even if the fits do not perfectly align with the galaxy. Approximately 30\% of the parent sample had successful ellipse fits in the F444W filter. The remaining $\sim$ 70\% of galaxies that failed ellipse fittings are overly disturbed, poorly resolved, and/or low surface brightness systems and removed from the sample. In addition, in a small number of cases, these were images of point sources conspicuously showing the JWST point spread function (PSF), in which secondary PSF features prevent a successful fit.
    \item[\textbf{Phase 2}] The ellipses fitted in the previous step do not have a specified centre, which may prevent the correct identification of highly inclined galaxies. We thus take the inner 40\% of the isophotes fitted to the galaxy in the first step, exclude the inner 10\% isophotes and take the average position of the centre of these isophotes as the galaxy centre. The choice for this range of radii ensures that one has enough pixels to compute a statistically robust position of the galaxy centre and simultaneously avoids strongly asymmetric structures, which are often at larger radii. By using isophotes all the way to 40\% of the fitted outer radius, we minimise the impact that a bright close-to-central point source would have in shifting the centre of isophotal fits from the correct galaxy centre. Visually inspecting the fits, we see that this is not a significant issue. It is also important to note that a bright point source overlapping with the galaxy will often result in the galaxy being visually unclassifiable, removing it from the sample at Phase 1. In addition, a function assesses the pixel values in a 10x10 pixels window centred at the new galaxy central pixel coordinates to ensure that bright foreground stars do not influence the determination of the galactic centre. To improve convergence stability against non-elliptical structures, including stars, \texttt{photutils.isophote} clears aberrant points from each isophote with a k-sigma algorithm. We then re-run \texttt{photutils.isophote} on F444W NIRCam galaxy images with fixed specified central positions. With fixed centres, the ellipse fits failed for approximately 26\% of the newly optimised sample. We verified that the failed ellipse fits correspond to galaxies with overly irregular or ambiguous morphology. These systems are also removed from the sample. We use the default scheme developed in \citet[][]{Jedrzejewski_1987}{}{} to decide when the fitting stops. By visually inspecting the fits, we verified that the fits only rarely stop before reaching the outskirts of the galaxy.
    \item[\textbf{Phase 3}] An inclination limit of $i \leq$ \ang{60} is applied to remove highly inclined galaxies as it is difficult to identify if a bar is present in these cases. We define the inclination of a galaxy by measuring the ellipticity of the outermost fitted ellipse,
    \begin{equation}
    e = 1 - \frac{b}{a} ,
    \end{equation}
    where $b$ is the minor axis length and $a$ is the major axis length. The inclination is defined as
    \begin{equation}
    \cos{i} = \frac{b}{a} .
    \end{equation}
    Approximately 34\% of the galaxies in the F444W filter of the newly optimised sample were seen to be too highly inclined and were removed from the sample. While highly-inclined, disturbed and edge-on galaxies have been removed in this and the previous phases of the optimisation, we note that a residual fraction of regular disc galaxies with inclination larger than $60^\circ$ still went through to the optimised sample despite this final optimisation step. Conservatively, we treated these galaxies as all remaining ones, which means our final bar fractions could be slightly underestimated (although, in some cases, a bar could be seen despite the high inclination). We speculate that a combination of the relatively poor physical spatial resolution and lower outer surface brightness compared to local galaxies is the cause behind these residual highly inclined galaxies remaining in the optimised sample.
\end{description}

\begin{table}
    \centering
    \begin{tabular}{ccccc}
    \hline
    \multirow{2}{*}{Phase} & \multicolumn{2}{c}{HST} & \multicolumn{2}{c}{JWST} \\\cline{2-5}
     & $N_{\rm {gal}}$ removed & $N_{\rm {gal}}$ remain & $N_{\rm {gal}}$ removed & $N_{\rm {gal}}$ remain\\
    \hline
    1 & 4980 & 465 & 3635 & 1583 \\
    2 & 230 & 235 & 416 & 1167 \\
    3 & 102 & 133 & 399 & 768 \\
    \hline
    \end{tabular}
    \caption{Number of galaxies removed from the sample and the resultant sample size after each optimisation phase. Col. 1: the optimisation phase. Col. 2 and Col. 3 are in the context of HST CANDELS F160W images. Col. 2: the number of galaxies which failed to meet the phase criteria. Col. 3: the sample size after the criteria are applied, with phase 1 being applied to the parent sample. Col. 4 and Col. 5 are the same as Col. 2 and Col. 3 but in the context of JWST CEERS F444W images.}
    \label{tab: Optimisation}
\end{table}

We applied this three-step optimisation procedure to our initial large CEERS F444W galaxy sample, ensuring elliptical isophotes could be fitted to the galaxy image with an identified galactic centre and that the galaxy was not edge-on. The resultant optimised sample of galaxies suitable to our analysis is 768 CEERS images in the NIRCam F444W filter (hereafter referred to as the optimised JWST sample). Of the optimised galaxy sample, 404 galaxies are between the redshifts $1 \leq z \leq 2$, and 364 galaxies are between the redshifts $2 < z \leq 3$. Before visual classifications, a co-author (DG) visually verified that all removed objects were indeed poorly resolved, overly faint/irregular, or too inclined. Table \ref{tab: Optimisation} gives the number of galaxies removed at each phase and the resultant galaxy sample size.

To measure the difference in the bar fraction between JWST and HST, we also applied the three-step optimisation procedure in our HST CANDELS F160W galaxy sample. This reduced our HST CANDELS sample to an optimised sample of 133 galaxies (hereafter referred to as the optimised HST sample). The reduced sensitivity and bluer wavelength range of HST means many of the galaxies are very pixelated, and features are difficult to discern. Therefore, the ellipse fitting technique failed on many of these galaxies, greatly reducing the optimised HST sample size.

\subsection{Disc identification}
\label{Sec: disc}

\begin{figure*}
    \centering
    \includegraphics[width=\textwidth]{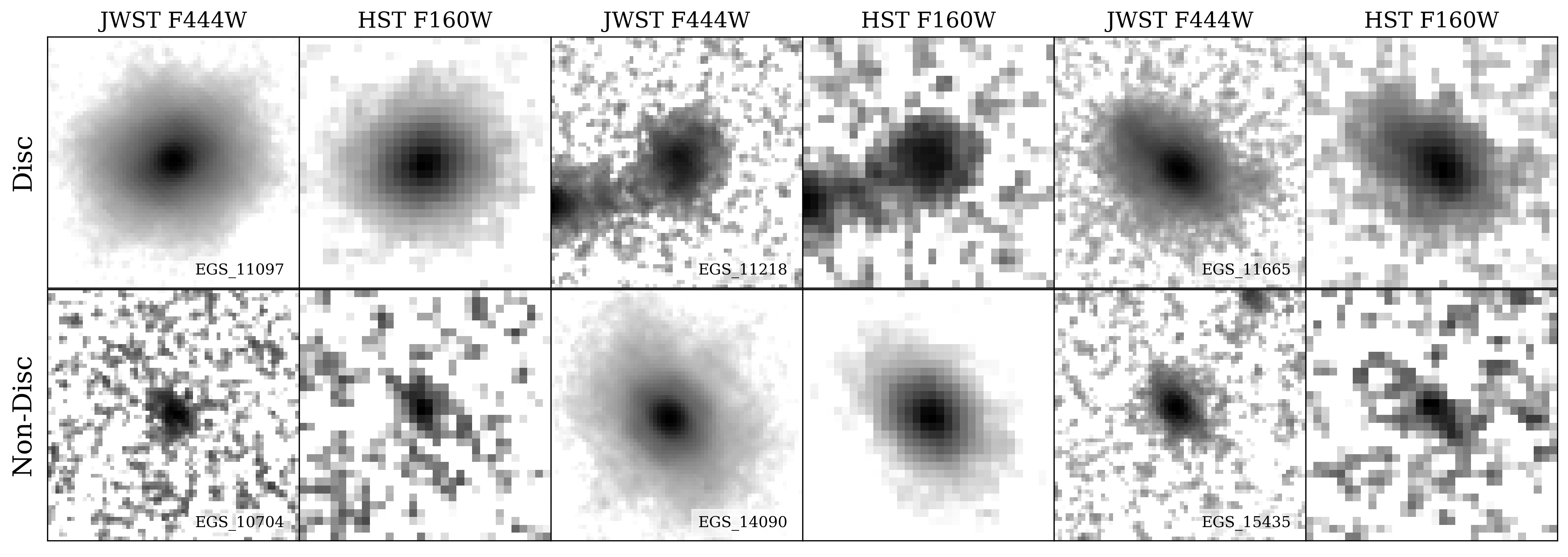}
    \caption{Rest-frame NIR logarithmic images of three disc (top row) and three non-disc (bottom row) galaxies. The three exemplars for each classification, with IDs in the lower right of the NIRCam F444W image, are shown in the JWST NIRCam F444W (left) and HST WFC3 F160W (right).}
    \label{Fig: disc_examples}
\end{figure*}

Naturally, the bar fraction is the number of barred galaxies divided by the number of disc galaxies, and we therefore need first to compute the latter. The variations in morphological appearance and dependence on orientation make disc galaxies challenging to identify visually. Given the different wavelength ranges probed by JWST and HST, we compute the disc fraction in our parent samples to avoid any potential bias from our optimisation procedure in the number of disc galaxies and extrapolate that to the optimised sample. This is also important as the optimised sample is significantly smaller than the parent sample, and uncertainties thus become more significant. We obtain the JWST disc fraction from the published visual classifications by co-author LF and collaborators \citep{Ferreira_2022}. Six independent participants visually classified $3559$ CEERS sources in their rest-frame NIR images, using the NIRCam filters F277W, F356W and F444W for the redshifts $z = 1.5 - 3.0$, which contained $1531$ discs. The disc fraction found by LF does not extend to the lower end of the redshift range selected for this study, however we assume that the disc fraction between redshifts 1 and 1.5 is the same as between redshifts 1.5 and 2. For the redshift bins $1 \leq z \leq 2$ and $2 < z \leq 3$, the disc fraction of the optimised JWST sample is thus $f_{disc} = 0.49 \pm 0.039$ and $f_{disc} = 0.39 \pm 0.046$, respectively. The systematic error on $f_{disc}$ is the standard error of $f_{disc}$ found by the six participants in \citet[][]{Ferreira_2022}{}{}. By extrapolating $f_{disc}$ to the optimised JWST sample, the number of disc galaxies $count_{disc} = 196$ and $143$ for $1 \leq z \leq 2$ and $2 < z \leq 3$, respectively.

For the HST sample, we use the visual classifications by \citet[][]{Kartaltepe_2015}{}{} to determine the optimised HST $f_{disc}$. For the redshift bins $1 \leq z \leq 2$ and $2 < z \leq 3$, the disc fraction of the HST parent sample is $f_{disc} = 0.75$ and $f_{disc} = 0.78$, respectively. It is important to point out that the detailed classification of \citet[][]{Kartaltepe_2015}{}{} includes disc galaxies in the category `irregular', which is also further subdivided in categories such as `spheroid and irregular' to separate disc galaxies that are irregular from both disc galaxies with regular morphology and spheroidal galaxies that are irregular. We thus include in $f_{disc}$ a smaller fraction of galaxies ($\sim15\%$) noted as irregulars (but after removing those noted as `spheroid and irregular') following strictly the classifications of Kartaltepe et al. It is important to include these galaxies in $f_{disc}$ because these are disc galaxies despite their disturbed morphology, and their discs can develop bars. The Large Magellanic Cloud is a notable example of an irregular galaxy with a disc that hosts a bar. Furthermore, \citet[][]{Kartaltepe_2015}{}{} show that the S\'ersic index and colour distributions of their galaxies classified as irregulars match the corresponding distributions of the galaxies classified as discs. And, indeed, some of the barred galaxies we find were classified as irregulars in \citet[][see our Table \ref{tab: bars}]{Kartaltepe_2015}{}{}. The systematic error on $f_{disc}$ is the standard error of $f_{disc} = \pm 0.026$ found by comparing the results from the three classifiers in \citet[][]{Kartaltepe_2015}{}{}. By extrapolating $f_{disc}$ to the optimised HST sample, $count_{disc} = 81$ and $19$ for $1 \leq z \leq 2$ and $2 < z \leq 3$, respectively. We note that the HST disc fraction may be a lower limit since \citet[][]{Nelson_2023}{}{}, using JWST, found that massive, dusty edge-on discs could have been missed as HST-dark galaxies. This would induce an overestimation of the bar fraction with HST data.


\begin{figure*}
    \centering
    \includegraphics[width=15 cm]{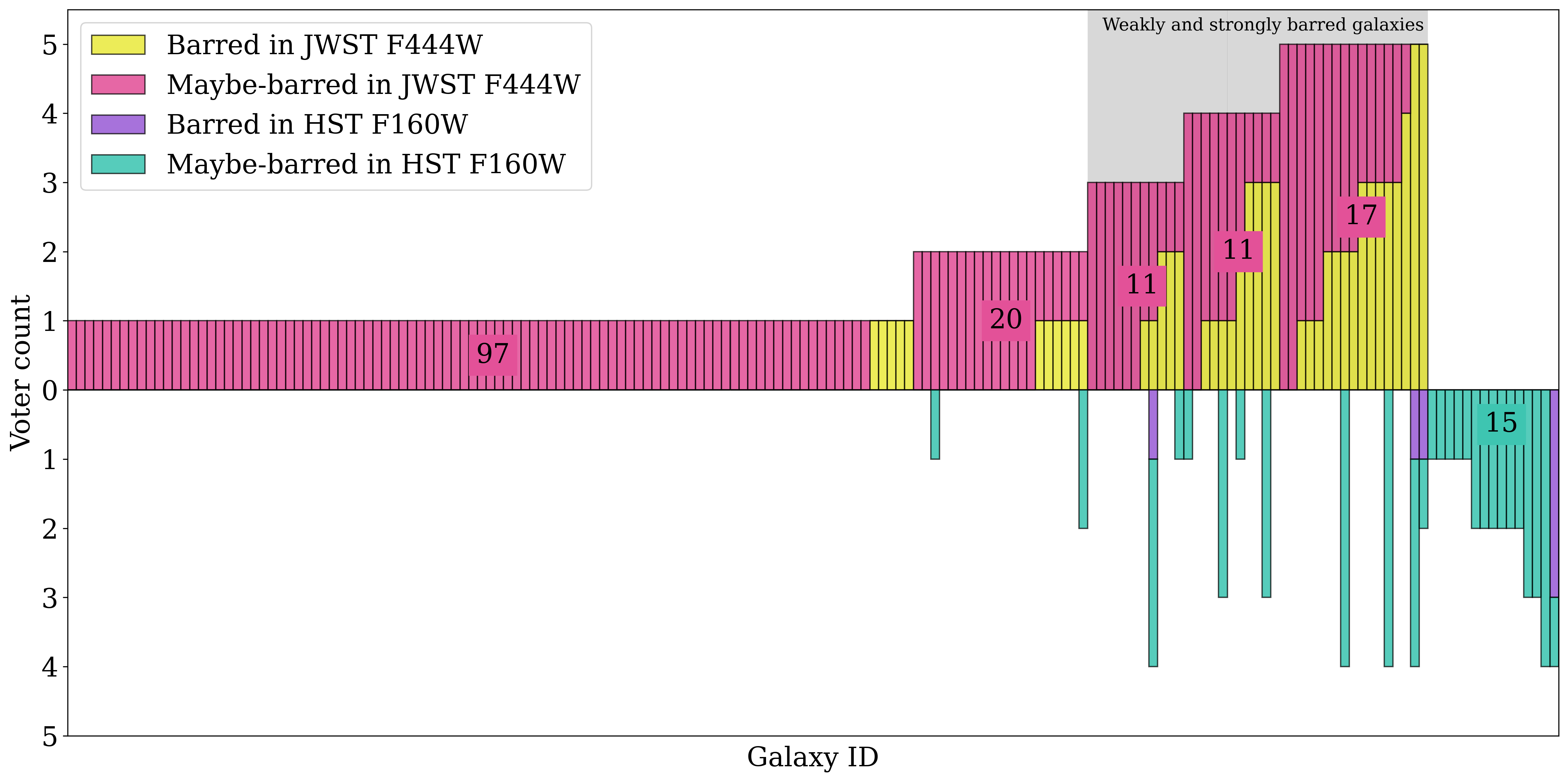}
    \caption{Distribution of the total number of barred in yellow (purple) and maybe-barred in pink (green) votes cast by the five participating co-authors to candidate galaxies in the optimised JWST (HST) sample. 171 galaxies received barred or maybe-barred votes in the NIRCam F444W filter or WFC3 F160W filter from the classifiers, and 597 galaxies received zero votes.} The number of galaxies in each voter count category is shown. A grey shaded area covers the galaxies which are classified as weakly or strongly barred. We exclude galaxies that received zero barred or maybe barred votes from this figure.
    \label{Fig: voter_hist}
\end{figure*}

Additionally, two co-authors (ZLC and DG) also visually classified the optimised JWST sample to study the stellar mass distributions of barred and unbarred galaxies in the optimised sample, discussed further below. The participants voted the galaxy to be a disc or non-disc based on the F444W NIRCam images and a log intensity radial profile. In principle, artefacts (discussed in \S~\ref{sec:reduction}) could mislead visual classifications, but these PSF effects are clearly distinguishable. The diffraction spikes mostly appeared as a large hexagon over the galaxy image, so the galaxy is not elongated in one direction preferentially. Therefore, we typically class these as non-discs/unidentifiable. To ensure we were not affected by less prominent artefacts, we checked for effects in the intensity radial profile of each galaxy. Figure \ref{Fig: disc_examples} shows three examples of disc and three non-disc galaxies in rest-frame JWST NIRCam F444W and HST WFC3 F160W filters. Non-disc galaxies can include strong PSF-affected sources, as shown by the central source in the figure. In this classification, the average disc fraction of the optimised JWST sample is $f_{disc} = 0.40 \pm 0.14$ for the full redshift range $1 \leq z \leq 3$ (the quoted uncertainty is the difference in $f_{disc}$ found by the two participants). The disc fraction derived in this study thus agrees with the disc fraction found by LF and collaborators \citep{Ferreira_2022}, which corresponds to $0.45 \pm 0.034$ when we employ their classifications and consider our optimised sample at $1.5 \leq z \leq 3$.

\subsection{Bar visual classifications}
\label{Sec: visual}

\begin{figure*}
    \centering
    \includegraphics[width=\textwidth]{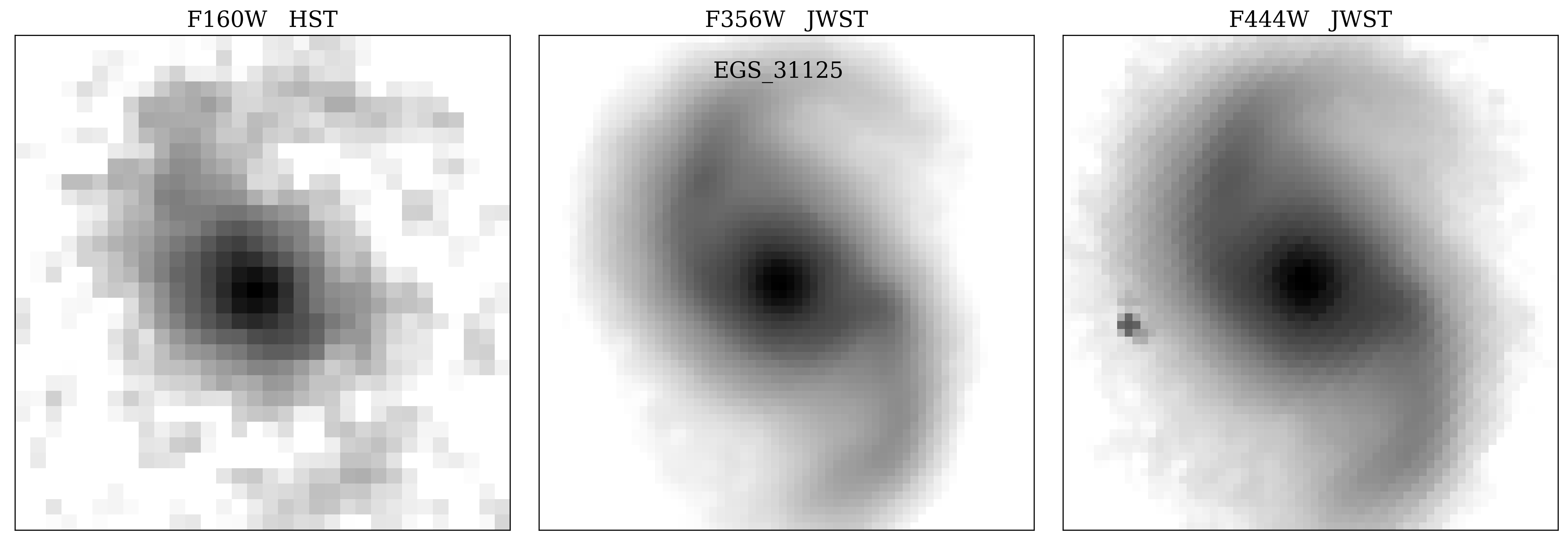}
    \caption{The logarithmic image of galaxy EGS$\_$31125 at redshift $z \simeq 2.06$, visually classified as strongly barred from the JWST NIRCam F444W image, shown in an HST filter and two JWST filters. From left to right: HST WFC3 F160W and JWST NIRCam F356W and F444W. This filter comparison demonstrates the effects of PSF, sensitivity and wavelength range on a galaxy image, particularly in the context of bars. The image shows EGS$\_$31125 in rest frame 0.52, 1.16, and 1.45 $\mu$m, respectively.}
    \label{Fig: filter_comp}
\end{figure*}

\begin{figure*}
    \centering
    \includegraphics[width=\textwidth]{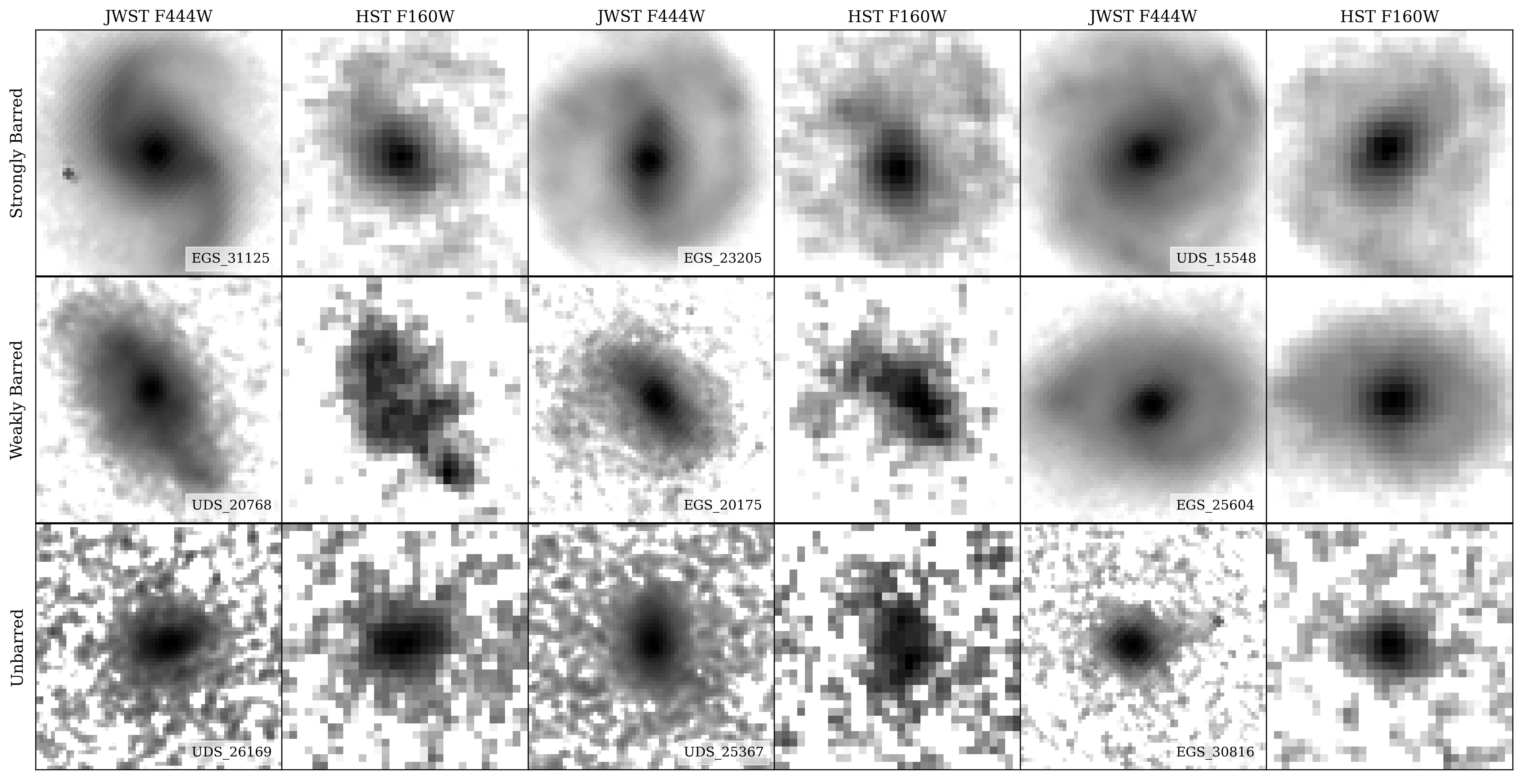}
    \caption{Rest-frame NIR logarithmic images of three strongly barred (top row), weakly barred (middle row) and unbarred (bottom row) galaxies. The three exemplars for each classification, with IDs in the lower right of the NIRCam F444W image, are shown in the JWST NIRCam F444W (left) and HST WFC3 F160W (right).}
    \label{Fig: bar_examples}
\end{figure*}

The optimised JWST sample was then visually classified by five co-authors (ZLC, DG, CdSF, TK and JN). The participants were asked to vote \textit{barred, maybe-barred} or \textit{unbarred} on the F444W NIRCam images. The votes were tallied, and a galaxy was classified as follows: a galaxy is classified as strongly barred if it obtained at least three out of five votes for barred; a galaxy is classified as weakly barred if it obtained two out of five votes for barred or at least three out of five votes for maybe-barred; a galaxy is classified as unbarred if it did not obtain the vote thresholds. Figure \ref{Fig: voter_hist} is a histogram of the number of barred and maybe-barred votes the co-authors gave on each galaxy in the optimised JWST sample. The figure does not show the galaxies where the co-authors were congruent about the galaxies being unbarred. The figure shows the difficulties in identifying bars as 75\% of the galaxies shown here are below the vote threshold, hence classified as unbarred.

The visual classification method was then repeated for the optimised JWST sample in the NIRCam F356W filter. The resolution marginally improves at this shorter wavelength, so structural features are better defined. However, this wavelength can be more subjected to dust extinction and star formation effects, so the bar-dominated evolved stellar populations may be only moderately traced. The overall bar fraction did not change between the two NIRCam filters, but a few weaker bars became stronger when observed in the filter F356W. Finally, the visual classification method was repeated again on the optimised HST sample. EGS\_31125 is shown in Figure \ref{Fig: filter_comp} in the three different filters employed: HST WFC3 F160W and JWST NIRCam F356W and F444W. EGS\_31125 is classified in the F444W filter as strongly barred and unbarred in the F160W filter. This figure clearly shows the impact of improved sensitivity and longer wavelength with JWST on the galaxy at redshift $z \simeq 2.06$ and how distinctive the disc structures become \citep[see also][]{Guo_2023}{}{}. On the other hand, it is interesting to note that 15 galaxies have been classified as barred in the HST sample but did not receive any vote when classified using the JWST data. These images were inspected again, and while these galaxies could indeed be barred, we note that in some cases, effects from the JWST PSF impact the classification. In other instances, details in the structure of the galaxy are better discerned in the JWST images, rendering the impression of a bar somewhat uncertain. We show examples of these galaxies in Figure \ref{Fig: HSTvotes} (Appendix \ref{App: B}).

\begin{table*}
    \centering
    \begin{tabular}{cccc|cc|c}
    \toprule
    \multicolumn{7}{c}{Sample Sizes} \\
    \midrule
    Sample & Redshift & $\rm{N}_{\rm {gal, HST}}$ & $\rm{N}_{\rm {gal, JWST}}$ & $\rm{N}_{\rm {gal, HST}}$ & $\rm{N}_{\rm {gal, JWST}}$ & Criteria applied \\
     & & & &\multicolumn{2}{c|}{Mass Complete} & \\
    \midrule
    Parent sample & $1 \leq z \leq 3$ & 5445 & 5218 & 1299 & 1180 & Redshift \\
    \midrule
    \multirow{3}{*}{Optimised sample} & $1 \leq z \leq 3$ & 133 & 768 & 126 & 368 & \multirow{3}{*}{ellipse fitting, $i \leq$ \ang{60}} \\
     & $1 \leq z \leq 2$ & 108 & 404 & 105 & 237 \\
     & $2 < z \leq 3$ & 25 & 364 & 21 & 131\\
    \midrule
    \multirow{3}{*}{Disc sample} & $1 \leq z \leq 3$ & 100 & 339 & 98 & 229 & \multirow{3}{18em}{HST discs from \citet[][]{Kartaltepe_2015}{}{} and JWST discs from \citet[][]{Ferreira_2022}{}{}} \\
     & $1 \leq z \leq 2$ & 81 & 196 & 81 & 157 & \\
     & $2 < z \leq 3$ & 19 & 143 & 17 & 72 & \\
    \midrule
    \multirow{3}{*}{Weakly Barred} & $1 \leq z \leq 3$ & 9 & 26 & 9 & 25 & \multirow{3}{*}{Visually classified bars} \\
     & $1 \leq z \leq 2$ & 8 & 21 & 8 & 20 & \\
     & $2 < z \leq 3$ & 1 & 5 & 1 & 5 & \\
    \midrule
    \multirow{3}{*}{Strongly Barred} & $1 \leq z \leq 3$ & 1 & 13 & 1 & 13 & \multirow{3}{*}{Visually classified bars} \\
     & $1 \leq z \leq 2$ & 1 & 8 & 1 & 8 & \\
     & $2 < z \leq 3$ & 0 & 5 & 0 & 5 & \\
    \midrule
    \multicolumn{7}{c}{The Bar Fraction} \\
    \midrule
    \multirow{3}{*}{Bar Fraction} & $1 \leq z \leq 2$ & $0.11^{+ 0.05}_{- 0.04}$ & $0.15^{+ 0.05}_{- 0.05}$ & $0.11^{+ 0.05}_{- 0.04}$ & $0.18^{+ 0.05}_{- 0.05}$ & \multirow{3}{*}{$\frac{\rm{N}_{\rm {weakly\_barred}} + \rm{N}_{\rm {strongly\_barred}}}{\rm{N}_{\rm {disc}}}$} \\
     & & & & & & \\
     & $2 < z \leq 3$ & $0.05^{+ 0.08}_{- 0.04}$ & $0.07^{+ 0.07}_{- 0.06}$ & $0.06^{+ 0.08}_{- 0.04}$ & $0.14^{+ 0.07}_{- 0.06}$ & \\
    \bottomrule
    \end{tabular}
    \caption{Progression of the galaxy sample sizes after the different selection and classification criteria are applied. Col. (1): the sample label. Col. (2): the redshift range. Col. (3): the number of galaxies after applying the criteria to HST CANDELS F160W images. Col. (4): the number of galaxies after applying the criteria to JWST CEERS F444W images. Col. (5) and (6): the same as col. (3) and (4) but with the 95\% mass completeness cut applied to the sample from \citet[][]{Duncan_2019}{}{}. Col. (7): the criteria applied. The bar fractions derived before and after applying the 95\% mass completeness limit are given for the two redshift bins at the bottom of the table. The bar fraction errors are explained in \S~\ref{Sec: fraction}.}
    \label{tab: Samples}
\end{table*}



\section{The bar fraction}
\label{Sec: fraction}

\begin{figure}
    \centering
    \includegraphics[width=\columnwidth]{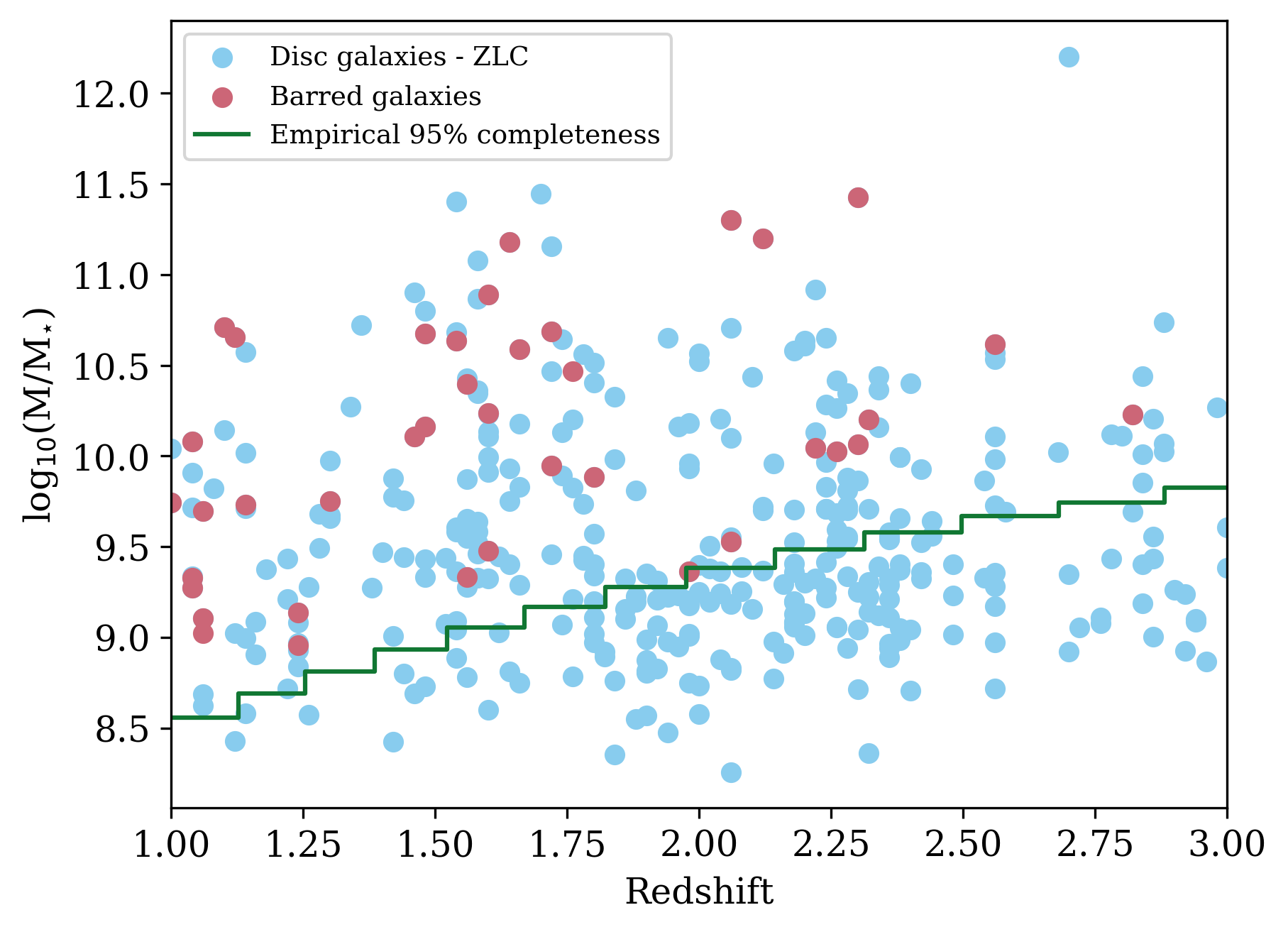}
    \caption{Distribution of stellar masses for the sample of disc galaxies as classified by ZLC in JWST CEERS between the redshifts $1 \leq z \leq 3$. Disc galaxies are shown in blue, while the weakly and strongly barred galaxies are in pink. A green step-wise line shows the 95\% empirical completeness of the sample \citep[see Figure 8 of ][]{Duncan_2019}{}{}. The parameter space below this line in this plot corresponds to a completeness fraction of $\approx85-90\%$.}
    \label{fig:mass}
\end{figure}

We aim to determine the fraction of the disc galaxy population at redshifts $z = 1 - 3$ hosting a bar. We visually classified the optimised JWST sample, which met the criteria described in \S~\ref{Sec: criterion}. The process is repeated for the optimised HST sample to explore if an increase in the bar fraction is found using JWST. Galaxies were classified as described in \S~\ref{Sec: visual}. Figure \ref{Fig: bar_examples} shows three examples of strongly barred, weakly barred and unbarred galaxies in the JWST NIRCam F444W and HST WFC3 F160W filters. The strongly barred galaxies have distinct stellar structures, while some weakly barred galaxies have less prominent outer discs. 

The bar fraction is found for two redshift bins, $1 \leq z \leq 2$ and $2 < z \leq 3$, to observe the evolution of the bar fraction. The redshift was only divided into two bins, as the number of barred galaxies is relatively small. In the optimised JWST sample, 29 galaxies were identified as barred in the lower redshift bin, where eight are strongly barred, and 21 are weakly barred, which decreased to ten barred galaxies in the higher redshift bin, where five are strongly barred, and five are weakly barred. All galaxies classified as barred are shown in Appendix \ref{App: A}: Figure \ref{Fig: strong} shows the strongly barred galaxies, while Figure \ref{Fig: weak} shows the weakly barred galaxies. The 39 barred galaxies are listed in Table \ref{tab: bars}, along with their photometric redshift and HST visual classifications from \citet{Kartaltepe_2015} of which only one galaxy was identified as barred. In the optimised HST sample, nine galaxies were identified as weakly or strongly barred in the lower redshift bin, and only one weakly barred galaxy was identified in the higher redshift bin.

\begin{figure*}
    \centering
    \includegraphics[width=\textwidth]{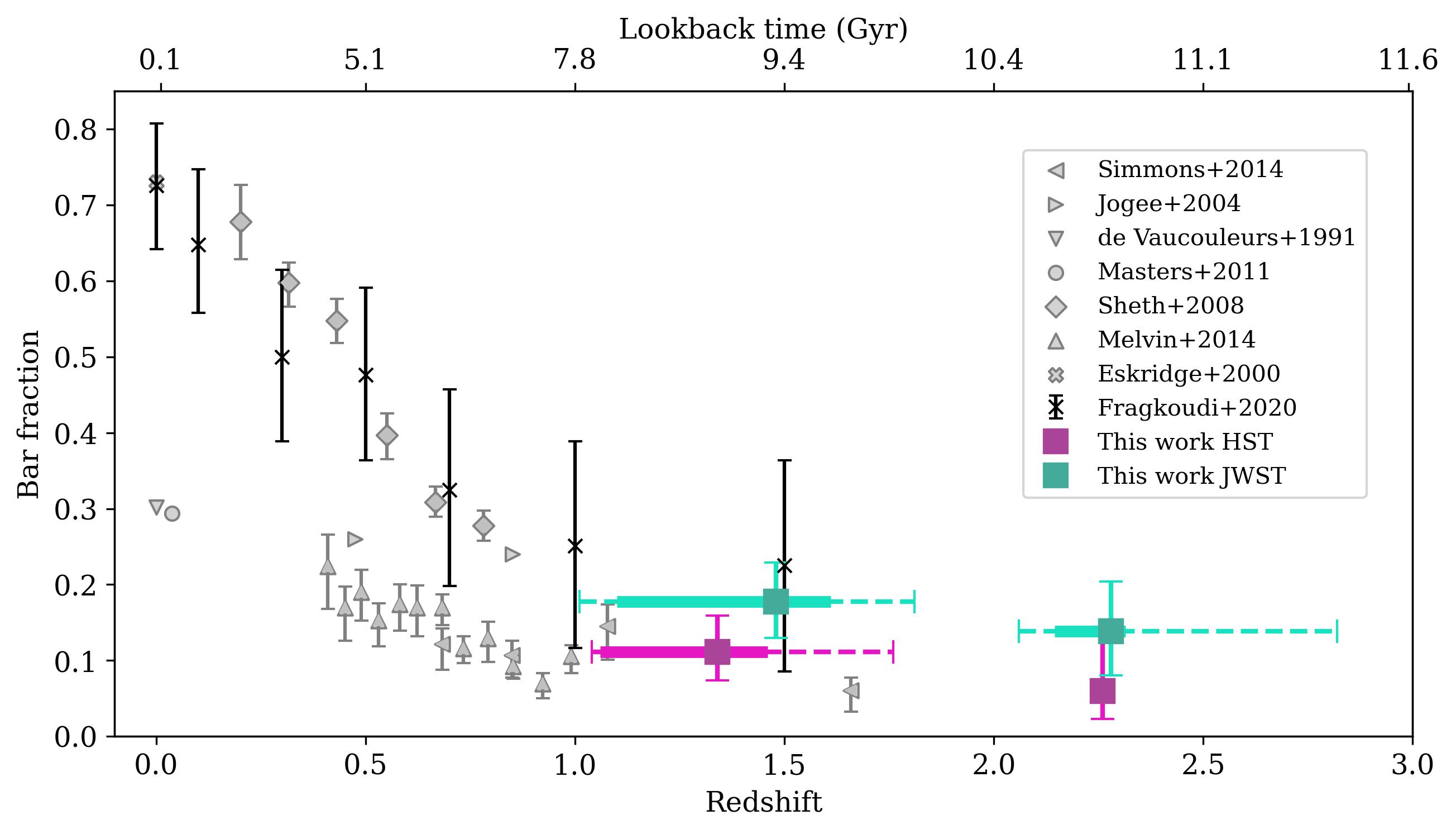}
    \caption{Evolution of the fraction of stellar bars in disc galaxies with redshift in the context of other bar assessment work using HST. The fractions of barred disc galaxies found in JWST NIRCam images are shown as green squares, and the fractions of barred disc galaxies found in this study in HST WFC3 images are shown as purple squares. The bar fraction was found for two redshift bins, $1 \leq z \leq 2$ and $2 < z \leq 3$, where the marker indicates the median redshift of the barred galaxies. All of the bar fraction errors, in this study, indicate the sum in quadrature of the systematic and statistical errors. A dashed line indicates the redshift range of barred galaxies. A thick solid line indicates the redshift range of the quartiles 25\%-75\% of the barred galaxies distribution. At low redshifts, \citet[][down-pointing triangle]{Vaucouleurs_1991}{}{} and \citet[][circle]{Masters_2011}{}{} found strong bars in a third of disc galaxies, while \citet[][cross]{Eskridge_2000}{}{} found strong and weak bars in over two-thirds of disc galaxies. \citet[][left-pointing triangles]{Simmons_2014}{}{}, \citet[][diamonds]{Sheth_2008}{}{} and \citet[][up-pointing triangles]{Melvin_2014}{}{} found a decreasing trend of the bar fraction for higher redshifts. \citet[][right-pointing triangles]{Jogee_2004}{}{} found a minimal decline in the bar fraction at higher redshifts. Finally, the bar fractions, as found in the Auriga cosmological simulations in \citet[][exes]{Fragkoudi_2020}{}{} are shown in black.}
    \label{Fig: bar_frac}
\end{figure*}

It is debated how the bar fraction depends on the stellar mass \citep[e.g.,][]{Barraza_2008,Sheth_2008,Nair_2010b,Masters_2011,Melvin_2014,Gavazzi_2015,Erwin_2018}{}{}. In Figure \ref{fig:mass}, we show the stellar mass distribution as a function of redshift for all disc galaxies in the optimised sample. The disc galaxies are taken from the classification of one of the participants in the disc classification procedure (ZLC). Still, we verified that qualitatively similar results are found regardless of the classifier. The 95\% empirical completeness limit of the sample, as estimated in \cite{Duncan_2019}, is indicated, showing that most of our sample is above or close to the completeness limit. Interestingly, this figure shows that barred galaxies tend to avoid the least massive galaxies at each redshift, in line with the results from \cite{Sheth_2008}. In order to calculate the bar fraction in a sample as complete in mass as possible, we apply the 95\% completeness step function as a mass limit, i.e., we select galaxies above the 95\% completeness line. This includes a new computation of the disc fractions as above but now only for galaxies above the 95\% completeness line, which slightly increases the disc fractions for both HST and JWST. For the redshift ranges $1 \leq z \leq 2$ and $2 < z \leq 3$ the JWST bar fraction is $\approx 17.8^{+ 5.1}_{- 4.8}$ per cent and $\approx 13.8^{+ 6.5}_{- 5.8}$ per cent, and the HST bar fraction is $\approx 11.2^{+ 4.7}_{- 3.8}$ per cent and $\approx 6.0^{+ 8.4}_{- 3.7}$ per cent, respectively. The uncertainties on these bar fractions are the sum in quadrature of the systematic and statistical errors, which are discussed below. For a 95\% complete stellar mass sample, the fraction of bars found in disc galaxies in JWST is approximately twice that in HST. Table \ref{tab: Samples} shows the progression of the JWST and HST galaxy sample sizes after the different selection and classification criteria are applied.


Figure \ref{Fig: bar_frac} shows the visually classified bar fraction versus redshift and lookback time in the context of other observational work assessing bar fractions using HST. The figure shows that previous results based on HST data indicate a decline in the bar fraction from lower to higher redshifts. While the JWST bar fraction also decreases from the redshift bin $1 \leq z \leq 2$ to the redshift bin $2 < z \leq 3$, the JWST bar fraction in the lower redshift bin is greater than the HST bar fraction in the same redshift bin. A dashed line indicates the redshift range of our visually identified barred galaxies, and a thick solid line indicates the distribution quartiles, i.e. 25\%-75\%. We identify the highest redshift strongly barred galaxy as EGS\_24268 at $z \simeq 2.32$ \citep[also found in][]{Guo_2023}{}{} and the highest redshift weakly barred galaxy as EGS\_22729 at $z \simeq 2.82$. 

The Jeffreys interval \citep[][]{Brown_2001,Gelman_2003}{}{} is used to determine the statistical uncertainty in the computed bar fractions. Considering the fraction of a large population with a given attribute (i.e., bars) and neither close to 0 nor 1, the Normal approximation can be assumed to derive uncertainties, but for small sample sizes and extreme population proportion values (e.g., the HST bar fraction at $2 < z \leq 3$), \cite{Cameron_2011} convincingly argues for a beta distribution quantile technique over the ‘normal approximation’ and the \citet[][]{Clopper_1934}{}{} approach, in which the Jeffreys ‘non-informative’ prior can be used interchangeably with the uniform prior. We adopt this method to estimate the full 68 per cent confidence intervals of the bar fraction in the two redshift bins. The sample used in this study is mass complete, meaning we do not account for incomplete sampling in the uncertainty estimates. On the other hand, the more important systematic errors in our analysis stem from the difficulty of defining a galaxy as a disc or barred galaxy. The fraction of disc galaxies was taken from \citet[][HST disc fraction]{Kartaltepe_2015}{}{} and \citet[][JWST disc fraction]{Ferreira_2022}{}{}. The standard error of the mean of the disc fraction found by the participants in the two independent studies is interpreted as the main systematic error in the bar fractions. Hence, we sum in quadrature the systematic and statistical errors of the bar fractions to obtain the final uncertainties quoted above. The statistical errors for the lower and higher redshift bins are, respectively, $^{+ 0.041}_{- 0.030}$ and $^{+ 0.083}_{- 0.035}$ for HST, and $^{+ 0.033}_{- 0.028}$ and $^{+ 0.046}_{- 0.035}$ for JWST. The systematic errors for the lower and higher redshift bins are, respectively, $\pm 0.023$ and $\pm 0.011$ for HST, and $\pm 0.039$ and $\pm 0.046$ for JWST.



\section{Discussion}
\label{Sec: discussion}

Using NIRCam F444W images, corresponding to NIR rest-frame at $1 \leq z \leq 3$, the visually identified fraction of disc galaxies hosting a bar at redshifts $z = 1 - 2$ is $\sim 18$ per cent, which decreases to $\sim 14$ per cent at redshifts $z = 2 - 3$. We found the bar fraction obtained from the F444W JWST images to be greater than that obtained using F160W HST images by a factor of about two, as shown in Figure \ref{Fig: bar_frac}.
Our value of the bar fraction at $z = 1 - 2$, which we derive using HST images, is in line with the estimate from \citet[][see their Fig. 6]{Simmons_2014}{}{} who also use HST data for their estimates. This begs the question, why do we find more bars in JWST than in HST images? Considering that the parent sample was chosen to contain sources present in both HST CANDELS and JWST CEERS and that the same bar-detection method was applied to both the JWST F444W and HST F160W images, the considerable difference between the JWST and HST bar fractions at each redshift bin implies that the identification of bars in disc galaxies is dependent on the sensitivity and wavelength range of the instrument; the bar fraction increases at longer wavelengths and with improved sensitivity. Spatial resolution does not play a role here. Defining the resolution of an instrument as the full-width half maximum (FWHM) of the empirical point spread function (PSF), the resolution of HST at 1.6 $\mu$m is \ang{;;0.151}\footnote{PSF FWHM taken from the HST user documentation: \url{https://hst-docs.stsci.edu/wfc3ihb/chapter-7-ir-imaging-with-wfc3/7-6-ir-optical-performance}} and the resolution of JWST at 4.44 $\mu$m is \ang{;;0.145}\footnote{PSF FWHM taken from the JWST user documentation: \url{https://jwst-docs.stsci.edu/jwst-near-infrared-camera/nircam-performance/nircam-point-spread-functions}}, and therefore, both instrumental setups have very similar resolution. 

Interestingly, the HST bar fractions at the two redshift bins do not change significantly after applying the 95\% mass completeness limit, whereas the JWST bar fraction at $2 < z \leq 3$ increases substantially (although within the error bars). This may reflect the ability of JWST to detect more low-mass discs at high redshifts than HST, but still not facilitate the finding of the presumably shorter bars in low-mass discs, since the spatial resolutions of the HST and JWST images used in this work are comparable.

Our results build upon the previous studies, which find that bar-driven internal evolutionary processes for settled disc populations begin at $z \simeq 1$, whereas our new results suggest this to be $z \simeq 2$ or more. Our study finds that a sizable population of barred galaxies exists at $z \leq 3$, implying that massive disc galaxies can become dynamically settled with prominent bars at a lookback time of $\sim 11$ Gyrs. The idea that bar-driven galaxy evolution happens in some cases at $z > 2$ is generally consistent with the early bar formation epochs estimated for local galaxies in the Time Inference with MUSE in Extragalactic Rings (TIMER) project \citep[][]{Gadotti_2019}{}{}. For NGC 4371, it has been estimated that the bar formation happened at $z \approx 2$ \citep[][]{Gadotti_2015}{}{}, while for NGC 1433, this happened at $z \approx 1$ \citep[][]{deSaFreitas_2023}{}{}. Nonetheless, it is important to point out that not necessarily all barred galaxies observed at $2 < z \leq 3$ will remain as a barred disc galaxy down to $z \approx 0$, as the galaxies in the TIMER sample: late violent mergers may destroy the bar, as well as the disc altogether.

In a recent study conducted by \cite{Guo_2023}, six strongly barred galaxies were identified at $z > 1$ using rest-frame NIR images from the first four pointings of CEERS. The six observed galaxies have a range in redshift from $z \approx 1.1$ to $z \approx 2.3$, using photometric redshifts \citep[see][]{Guo_2023,Stefanon_2017}{}{}. In a cross-check, we find that all barred galaxies identified by Guo et al. were also classified by us as barred.

Several previous studies have found a decline in the fraction of bars in disc and spiral galaxies with redshift, however mass- and volume-limits vary between the studies, along with the bar classification method. \cite{Sheth_2008} observe the evolution of the bar fraction at redshifts $0.2 < z < 0.84$ from luminous (brighter than $L^{\star}_{V}$) face-on spiral galaxies in the COSMOS 2 deg$^{2}$ field. The classification methods used in Sheth et al. are ellipse-fitting and visual, which are cross-checked, and an agreement of 85\% is found. \cite{Masters_2011} found the bar fraction of a volume-limited visually selected SDSS sample using Galaxy Zoo at redshifts $0.01 < z < 0.06$ and $M_{r} < -19.38$. \cite{Melvin_2014} use visually selected galaxies via Galaxy Zoo from COSMOS HST images at redshifts $0.4 \leq z \leq 1.0$ with an applied stellar mass limit of log(M$_{\star}$/M$_{\odot}$) $\geq 10$. The bar fraction was extended to redshifts $0.5 \leq z \leq 2.0$ in \cite{Simmons_2014} through the visually selected CANDELS galaxies via Galaxy Zoo with an absolute $H-$band magnitude limit of $H < 25.5$. With the work of Simmons et al. overlapping with the lower redshift bin of our study and using visually identified CANDELS galaxies, we found that our results are in good agreement. Although many studies have found a decrease in the bar fraction at $z = 0 - 1$, some find little or no evolution of the bar fraction. \cite{Jogee_2004} identified bars in spiral galaxies using three independent techniques and found the fraction of bars to be $\sim 30 \pm 6$ per cent in COSMOS-ACS galaxies at redshifts $z \sim 0.2 - 0.7$ and $z \sim 0.7 - 1.0$, with completeness cuts of $M_{V} \leq -19.3$ and -20.6, respectively. \cite{Elmegreen_2004} also found a constant bar fraction of $\sim 23 \pm 3$ per cent at redshifts $z \sim 0.0 - 1.0$ in COSMOS-ACS galaxies.

A direct comparison between the results from these various studies is difficult to accomplish given the different techniques employed to identify bars and the different sample selection criteria. In particular, \cite{Erwin_2018} shows that in the local Universe the bar fraction depends strongly on galaxy mass, with a peak at $\rm {M}_{\star} \sim 10^{9.7} \rm {M}_{\odot}$, declining towards both higher and lower masses. At redshifts $0.2 \leq z \leq 0.6$ for a mass complete sample of $\rm M > 10^{10.5} \rm M_{\odot}$ galaxies in the COSMOS field, \cite{Cameron_2010} found the bar fraction of early-type discs with intermediate stellar masses to be twice that of late-type discs, and is reversed for high stellar masses. In this context, it is important to highlight that our sample probes the galaxy population with masses above $\approx 10^9 \rm {M}_{\odot}$, which at redshift zero may reflect the peak in the bar fraction distribution. Considering all barred galaxies we find in our study, their mean stellar mass is $\rm {M}_{\star} \sim 1.2 \times 10^{10} \rm{M}_{\odot}$, with a standard deviation of $\sim 5.8 \times 10^{10} \rm {M}_{\odot}$.

With the different redshift ranges and stellar masses probed, as well as rest-frame wavelength ranges, samples and techniques employed to find bars and disc galaxies in the different studies, it is clear that while Fig. \ref{Fig: bar_frac} presents an interesting summary of the findings from different studies, a direct comparison between these studies must account for a number of effects. One could venture into accounting, for example, for the various mass ranges by assuming a $z=0$ variation of the bar fraction with mass and translating it to the samples probed at higher redshifts. However, this exercise would have to assume that such variation is constant with redshift, an assertion that has not yet been investigated with enough depth to the best of our knowledge. Moreover, the other effects mentioned may be as important, and recipes to account for those are not straightforward to devise.

Using the magnetic-hydrodynamical cosmological simulation TNG50 \citep[][]{Nelson_2019}{}{}, \cite{Rosas_Guevara_2022} found that $\rm M_{\star} \geq 10^{10} \rm M_{\odot}$ spiral galaxies with bar formation are present as early as $z = 4$. When an angular resolution limit of twice the HST $I$-band angular PSF FWHM was applied, the fraction of bars dropped to a tenth of its original value at $z = 2$, reconciling theoretical predictions and observations. The rapid onset of bar formation in massive galaxies at early cosmic times has been predicted in numerical simulations for baryon-dominated systems \citep[see e.g.,][]{Algorry_2017,Fragkoudi_2021,Bland_2023}{}{}. And, in fact, recent studies have found candidate barred galaxies at higher redshifts than our limit at $z = 3$. \citet[][]{Costantin_2023}{}{} used HST and JWST images in multiple filters to study a galaxy at $z \simeq 3$ in the mass range of Milky Way progenitors and found evidence of a bar. Going further in wavelength than what is possible with JWST, \citet[][]{Tsukui_2023}{}{} and \citet[][]{Smail_2023}{}{} explored the sub-millimetre domain to find dusty, star-forming galaxies at $z > 4$ with morphology indicative of a bar. If these galaxies are confirmed as barred, they provide strong constraints to cosmological simulations.

Some of the previous observational studies discussed above suggest that the decrease in the bar fraction in massive disc galaxies out to $z \sim 1$ could be due to minor merger events that keep the disc dynamically hot. However, depending on the details of the merger/flyby interaction, this could, in fact, tidally induce bar formation \citep[e.g.,][]{Berentzen_2003,Peschken_2019}{}{}.

The decline in the bar fraction in disc galaxies could be explained as a result of the decreasing physical spatial resolution with redshift. The ellipticity of bars at poorer resolution decreases, leading to a rounder, less elongated and compact bar, making the stellar bar less distinguishable. The perceptibility of a bar could be considerably affected by a clumpy outer disc, a bright central bulge and/or the angular size of the bar \citep[e.g.,][]{Lee_2019}{}{}. In the context of our results using JWST, the PSF FWHM for the JWST F444W filter is \ang{;;0.145}. The median redshift for barred galaxies between $1 \leq z \leq 2$ is $z = 1.48$, corresponding to a mean linear resolution of $\approx 1.26$ kpc. As for the redshift bin $2 < z \leq 3$, the median redshift of barred galaxies is $z = 2.28$, corresponding to a mean linear resolution of $\approx 1.22$ kpc. Bars smaller in angular size could have been preferentially missed at the high redshifts explored in this study. In a volume-limited SDSS galaxy sample where bars were identified through ellipse fits and Fourier analysis, \cite{Aguerri_2009} established that only bars with lengths above 2.5 times the FWHM can be identified. The proposal that the high-redshift bar fraction is systematically underestimated was thoroughly discussed in the context of a mass- and volume-limited S$^{4}$G galaxy sample in \cite{Erwin_2018}, where visual bar length measurements were obtained from \cite{Herrera_2015}. Erwin successfully reproduced SDSS bar fraction trends using SDSS observational parameters in simulations on the S$^{4}$G galaxy sample and suggested a bar length detection limit of $\sim 2$ times the FWHM. Applying these detection limits on NIRCam F444W images implies that bars shorter than $\sim 2.5 - 3$ kpc in radius (semi-major axis) are missed in our study.

Our resolution limit thus indicates that all bars we detect in this study are longer than $\approx 3$ kpc. 
In fact, \citet[][]{Erwin_2005}{}{} found that the mean bar semi-major axis is 3.3 kpc for early-type disc galaxies and 1.5 kpc for late-type disc galaxies \citep[see also][]{Gadotti_2011}{}{}. Therefore, unless the bar size distribution at high redshifts differs from the local distribution, even with JWST, we are likely missing a sizeable fraction of barred galaxies \citep[see also the discussion in ][]{Liang_2023}{}{}. In a sample of massive galaxies ($\rm M_{\star} \geq 10^{10} \rm M_{\odot}$, $0.02 \leq z \leq 0.07$) studied in \cite{Gadotti_2011}, there are not many bars that are shorter than 3 kpc (see his Fig. 1) although the author points out that due to resolution limits, he may also miss bars with semi-major axis below $2 - 3$ kpc. However, in \cite{Erwin_2005}, mass is not presented, so a direct comparison is not straightforward. \cite{Erwin_2019}, on the other hand, shows that bar length increases with mass for galaxies more massive than log ($\rm M_{\star} / \rm M_{\odot}$) $\geq 10.1$ for local galaxies, and a substantial fraction of the galaxies in his study has bars shorter than 3 kpc. 

Not only absolute bar length but the ratio of bar length to the galaxy size (e.g., disk scale length $h$, or parameters such as $R_{50}$ or $R_{90}$) may be more useful to compare at different redshifts, since it has been shown that the galaxy size also evolves \citep[][mostly for massive early-type galaxies but also in the case of
disk galaxies]{Trujillo_2007,Buitrago_2008,Van_der_Wel_2014,Buitrago_2014,Whitney_2019}{}{}. \cite{Kim_2021} measured bar length for galaxies at $0.2 \leq z \leq 0.84$ and found that the mean length of the bar is $\sim 5$ kpc for galaxies with log($\rm M_{\star}/\rm M_{\odot}$) $\geq 10$ (see their Fig. 2). However, the normalised bar length $R_{bar} / h$ of galaxies at $0.2 \leq z \leq 0.84$ in the study of \cite{Kim_2021} is similar to that of local bars in \cite{Gadotti_2011}. We postpone a thorough discussion on these aspects to a future paper, in which we will also present measurements of the bar length and its evolution at higher redshifts. 

It is also interesting to note that the abundance of weakly barred galaxies significantly declines at the higher redshift bin, more so than the abundance of strongly barred galaxies. This suggests the presence of another possible observational bias. While the linear resolution remains similar between $1\leq z\leq2$ and $2\leq z\leq3$ (as discussed above), cosmological surface brightness dimming is significantly more powerful at the higher redshift bin. This could diminish our ability to see weaker bars, particularly at the higher redshift bin, and this bias could produce a relative drop in the total bar fraction at the higher redshift bin even if the bar fractions in the two redshift bins are in reality comparable.



\section{Summary and conclusions}
\label{Sec: conclusion}

To derive the fraction of stellar bars in disc galaxies at high redshifts is an essential step towards understanding the onset of bar-driven galaxy evolution, which was found in previous studies using rest-frame optical HST images to occur at $z \sim 1$. However, stellar bars are populated by evolved stars emitting strongly at longer wavelengths, and thus, bars can be more effectively identified in rest-frame NIR images. 

In this study, we observe the evolution of the bar fraction at redshifts $z = 1 - 3$ in a sample of galaxies present in both HST CANDELS and JWST CEERS and PRIMER and compare the results obtained after using rest-frame optical HST images and rest-frame NIR JWST images for galaxies in the same parent sample. We use the longest-wavelength JWST NIRCam F444W filter to trace the underlying stellar mass distribution as best as possible. The initial parent sample of 1180 galaxies is optimised to produce a sample in which bars can be more robustly identified, in particular by removing galaxies with peculiar morphology and galaxies in a close to edge-on projection, with an inclination limit of $i \leq$ \ang{60}. After optimisation, the parent sample is reduced to 368 galaxies in the JWST F444W filter and 126 galaxies in the HST F160W filter. Five co-authors visually classified all galaxies in the two optimised samples, searching for bars supported by radial profiles of isophotal ellipticity and position angle.

To observe the evolution of stellar bars in disc galaxies, we used published disc classifications from \citet[][JWST disc classifications]{Ferreira_2022}{}{} and \citet[][HST disc classifications]{Kartaltepe_2015}{}{}. The fraction of bars in disc galaxies was thus derived for two redshift bins, $1 \leq z \leq 2$ and $2 < z \leq 3$, with robust photometric redshifts and ensuring a mass completeness above 95\%. The bar fractions we found in JWST F444W are, respectively, $\approx 17.8^{+ 5.1}_{- 4.8}$ per cent and $\approx 13.8^{+ 6.5}_{- 5.8}$ per cent for the lower and higher redshift bins. In HST F160W, we found the bar fractions to be $\approx 11.2^{+ 4.7}_{- 3.8}$ per cent and $\approx 6.0^{+ 8.4}_{- 3.7}$ per cent for the lower and higher redshift bins, respectively. Notably, at lower redshifts, we find 20 more barred galaxies in JWST, which were not identified in HST images and at higher redshifts, we classify nine more galaxies as barred. We thus found the bar fraction to be approximately two times greater in JWST F444W than in HST F160W, hence the restframe NIR bar fractions are twice the optical, showing that the detectability of stellar bars depends significantly on the wavelength range and the sensitivity of the instrument. A decrease in the bar fraction is observed at higher redshifts, but the trend could be due to shorter bars being preferentially missed from this study. We detect a substantial number of barred galaxies at redshifts $z \leq 3$, implying that bar-driven galaxy evolution could commence at least in some galaxies at a lookback time $\sim 11$ Gyrs, given that some bar-driven processes, such as promoting gas inflow along the bar leading edges, are thought to proceed quickly after bar formation \citep[$\sim 0.1$ Gyrs; see, e.g.,][]{Athanassoula_1992, Seo_2019, Baba_2020}{}{}. In fact, \cite{Guo_2023} have recently reported the finding of a barred galaxy at $z \approx 2.3$, and other teams have reported candidate barred galaxies beyond redshift three \citep[][Amvrosiadis et al. (2023), subm.]{Costantin_2023}{}{} and even beyond redshift four \citep[][]{Tsukui_2023, Smail_2023}{}{}. In this study, the highest redshift strongly and weakly barred galaxies found are at $z \simeq 2.3$ and $z \simeq 2.8$, respectively.

This study does not extend beyond $z = 3$ to remain in the rest-frame NIR and better detect the evolved stellar populations within the bar. Interesting investigations can be done on the bar fraction dependence on galaxy stellar mass and the evolution of the bar length, which are beyond the scope of this paper but will be explored in future papers. This study used the first four pointing of CEERS, and a future paper will present an enlarged census of the bar fraction at redshifts $1 \leq z \leq 3$ using the remaining six CEERS pointings.


\section*{Acknowledgements}

We thank the anonymous referee for a constructive and timely report. We also thank Peter Erwin for sending useful comments. ZLC acknowledges funding from the Science and Technology Facilities Council ST/X508354/1. This work was supported by STFC grants ST/T000244/1 and ST/X001075/1. TK acknowledges support from the Basic Science Research Program through the National Research Foundation of Korea (NRF), funded by the Ministry of Education (RS-2023-00240212 and No. 2019R1I1A3A02062242) and the grant funded by the Korean government (MSIT) (No. 2022R1A4A3031306 and WISET 2022-804). JN acknowledges funding from the European Research Council (ERC) under the European Union’s Horizon 2020 research and innovation programme (grant agreement No. 694343). EA thanks the CNES for financial support. For the purpose of open access, the authors have applied a Creative Commons Attribution (CC BY) to any Author Accepted Manuscript version arising.

\section*{Data Availability}

This work used Astropy \citep[][]{Astropy_2013}{}{} and PHOTUTILS \citep[][]{Bradley_2022}{}{}. The specific observations analyzed can be accessed via \url{https://doi.org/10.17909/xm8m-tt59}, and the visual classifications from \cite{Ferreira_2022} are publicly available at \url{https://github.com/astroferreira/CEERS_EPOCHS_MORPHO/}



\bibliographystyle{mnras}
\bibliography{Le_Conte_references}



\appendix
\onecolumn
\newpage
\section{Strongly and weakly barred galaxies}
\label{App: A}

\begin{figure}
    \centering
    \includegraphics[width=\textwidth]{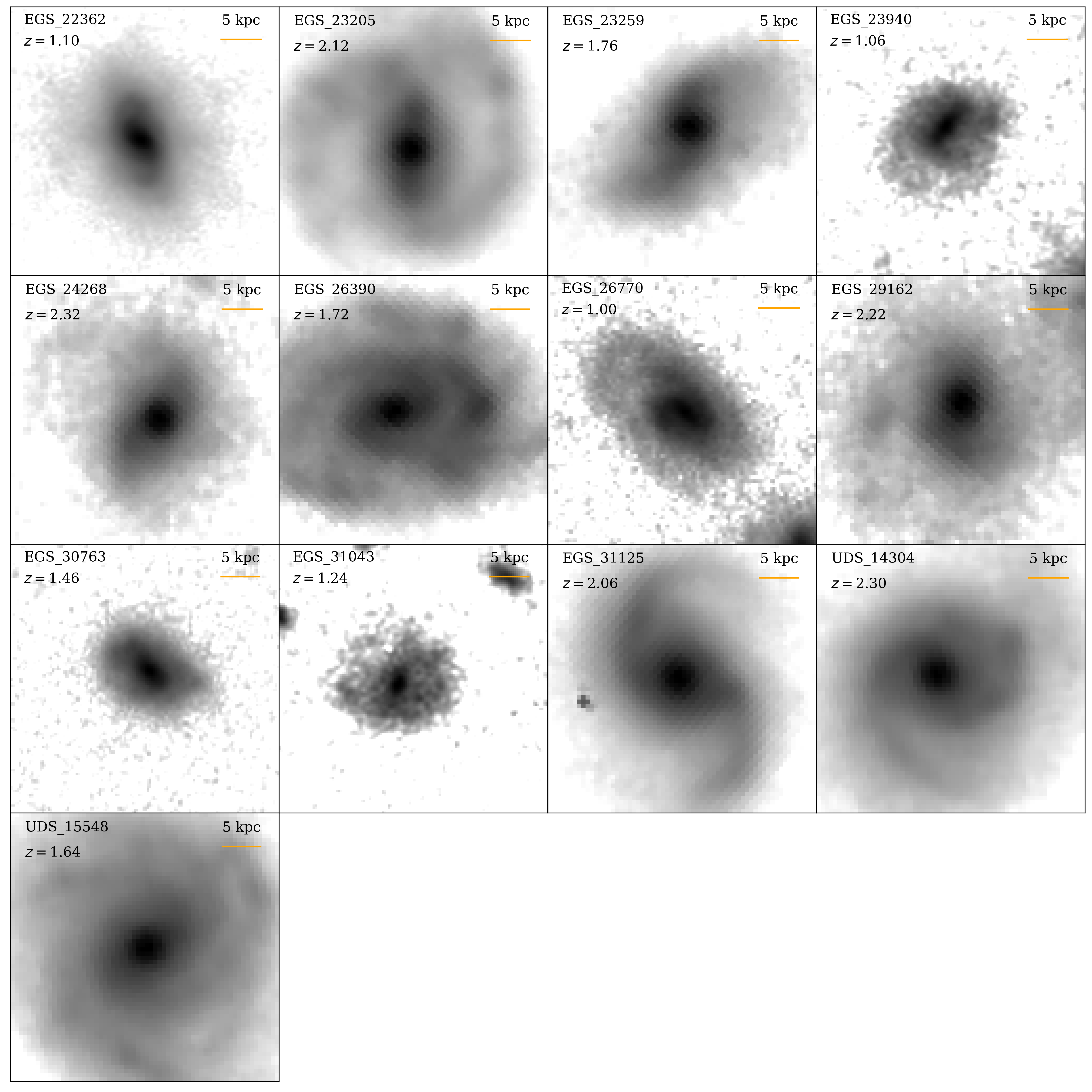}
    \caption{Rest-frame NIR logarithmic images of strongly barred galaxies using the JWST NIRCam F444W filter between the redshifts $1 \leq z \leq 3$. The redshift of the galaxy is noted in the upper left corner of each image. A 5 kpc scale is given in the upper right corner of each image \citep[calculated using][]{Wright_2006}{}{}.}
    \label{Fig: strong}
\end{figure}

\begin{figure}
    \centering
    \includegraphics[width=\textwidth]{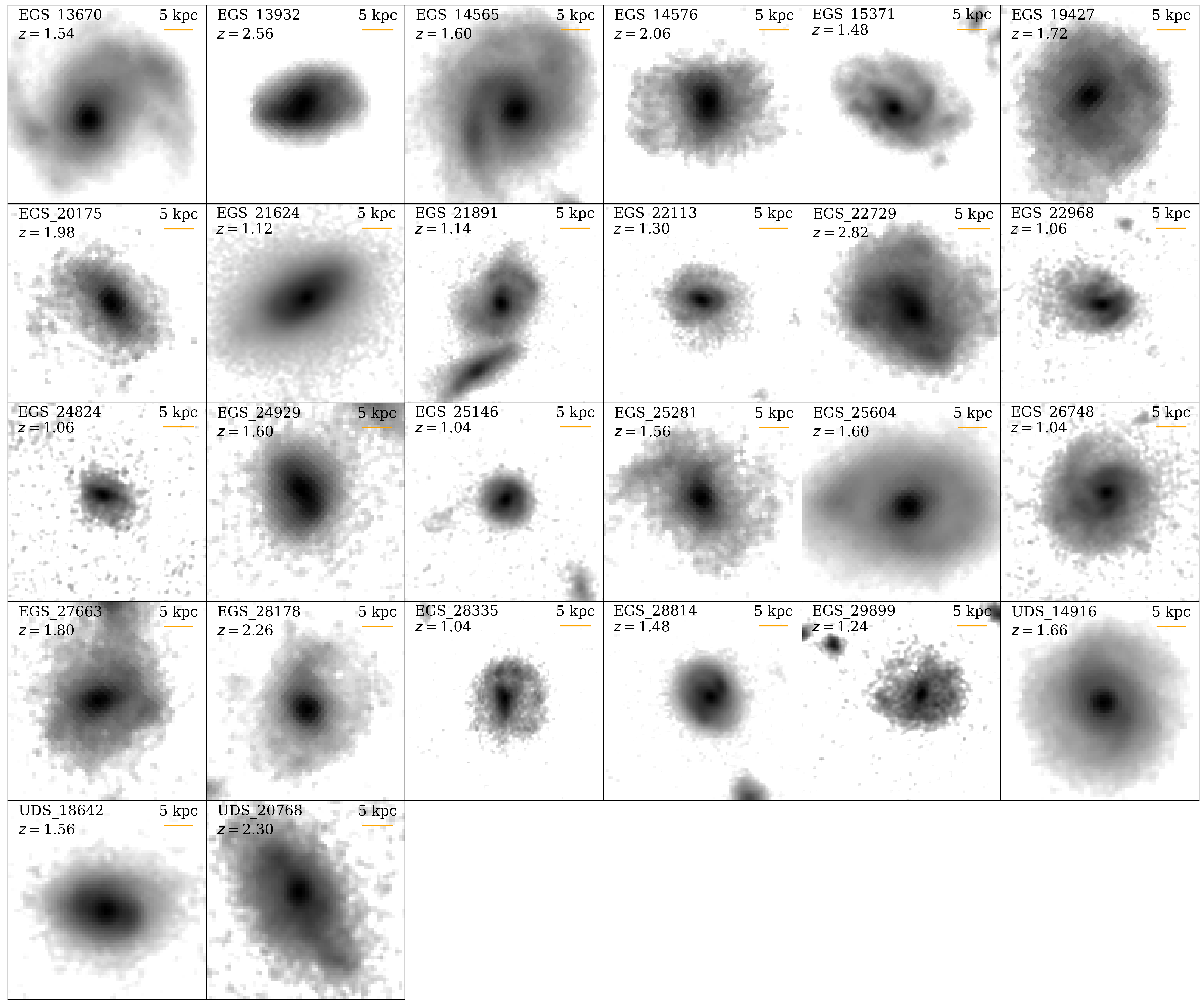}
    \caption{Rest-frame NIR logarithmic images of weakly barred galaxies using the JWST NIRCam F444W filter between the redshifts $1 \leq z \leq 3$. The redshift of the galaxy is noted in the upper left corner of each image. A 5 kpc scale is given in the upper right corner of each image \citep[calculated using][]{Wright_2006}{}{}.}
    \label{Fig: weak}
\end{figure}

In this study, we have visually identified CEERS galaxies as strongly barred, weakly barred or unbarred in JWST NIRCam F444W images between redshifts $1 \leq z \leq 3$. Figure \ref{Fig: strong} shows the strongly barred galaxies, while Figure \ref{Fig: weak} shows the weakly barred galaxies identified in this study. Table \ref{tab: bars} lists the HST classifications from \citet[][]{Kartaltepe_2015}{}{} of all barred galaxies, along with their photometric redshifts and stellar masses from \citet[][]{Duncan_2019}{}{}.

\begin{table*}
    \centering
    \begin{tabular}{ccccl}
    \hline
    Galaxy ID  & $z$  & log$_{10}$(M$_{\star}$/M$_{\odot}$)  & Bar Type  & Class \\
    \hline
    EGS\_13670 & 1.54 & 10.63 & Weak & Face-on barred disc with spiral arms \\
    EGS\_13932 & 2.56 & 10.61 & Weak & Disc \\
    EGS\_14565 & 1.60 & 10.24 & Weak & Barred disc dominated with spiral arms \\
    EGS\_14576 & 2.06 & 9.53 & Weak & Disc \\
    EGS\_15371 & 1.48 & 10.67 & Weak & Asymmetric disc \\
    EGS\_19427 & 1.72 & 9.95 & Weak & Disc with non-interacting companion \\
    EGS\_20175 & 1.98 & 9.36 & Weak & Disc \\
    EGS\_21624 & 1.12 & 10.65 & Weak & Disc dominated spheroid \\
    EGS\_21891 & 1.14 & 9.73 & Weak & Interacting disc with spiral arms \\
    EGS\_22113 & 1.30 & 9.75 & Weak & Disc \\
    EGS\_22362 & 1.10 & 10.71 & Strong & Spheroidal disc \\
    EGS\_22729 & 2.82 & 10.23 & Weak & Irregular asymmetric disc \\
    EGS\_22968 & 1.06 & 9.11 & Weak & Disc with spiral arms \\
    EGS\_23205 & 2.12 & 11.20 & Strong & Disc \\
    EGS\_23259 & 1.76 & 10.47 & Strong & Interacting asymmetric irregular disc \\
    EGS\_23940 & 1.06 & 9.70  & Strong & Disc with spiral arms and non-interacting companion \\
    EGS\_24268 & 2.32 & 10.20 & Strong & Asymmetric irregular \\
    EGS\_24824 & 1.06 & 9.03 & Weak & Disc \\
    EGS\_24929 & 1.60 & 9.48 & Weak & Asymmetric irregular with non-interacting companion \\
    EGS\_25146 & 1.04 & 9.27 & Weak & Disc with non-interacting companion \\
    EGS\_25281 & 1.56 & 9.33 & Weak & Face-on disc dominated spheroid \\
    EGS\_25604 & 1.60 & 10.89 & Weak & Face-on spheroidal disc \\
    EGS\_26390 & 1.72 & 10.69 & Strong & Face-on asymmetric disc \\
    EGS\_26748 & 1.04 & 10.08 & Weak & Face-on disc \\
    EGS\_26770 & 1.00 & 9.74  & Strong & Interacting face-on disc \\
    EGS\_27663 & 1.80 & 9.88 & Weak & Face-on disc with non-interacting companion \\
    EGS\_28178 & 2.26 & 10.03 & Weak & Face-on asymmetric disc dominated irregular \\
    EGS\_28335 & 1.04 & 9.33 & Weak & Disc with spiral arms \\
    EGS\_28814 & 1.48 & 10.16 & Weak & Asymmetric irregular disc \\
    EGS\_29162 & 2.22 & 10.04 & Strong & Interacting asymmetric irregular \\
    EGS\_29899 & 1.24 & 8.96 & Weak & No classification found \\
    EGS\_30763 & 1.46 & 10.11 & Strong & Asymmetric irregular \\    
    EGS\_31043 & 1.24 & 9.14  & Strong & Face-on disc with non-interacting companion \\
    EGS\_31125 & 2.06 & 11.30 & Strong & Face-on disc with spiral arms \\
    UDS\_14304 & 2.30 & 11.43 & Strong & Interacting spheroidal point source \\
    UDS\_14916 & 1.66 & 10.59 & Weak & Face-on disc dominated spheroid \\
    UDS\_15548 & 1.64 & 11.18 & Strong & Edge-on disc with non-interacting companion \\
    UDS\_18642 & 1.56 & 10.40 & Weak & Asymmetric disc \\
    UDS\_20768 & 2.30 & 10.07 & Weak & Disc \\
    \hline
    \end{tabular}
    \caption{The 39 visually classified strongly and weakly barred galaxies. Col. (1): The ID number is from the original SExtractor catalogue based on the full EGS and UDS mosaic. Col. (2): CANDELS-based catalogue photometric redshifts \citep{Duncan_2019}. Col. (3): CANDELS-based catalogue stellar masses \citep{Duncan_2019}. Col. (4): The classification of bar strength determined from JWST NIRCam images. Col. (5): HST visual classifications from \citet{Kartaltepe_2015}.}
    \label{tab: bars}
\end{table*}

\newpage
\section{Impact of instrument sensitivity on classifications}
\label{App: B}

\begin{figure}
    \centering
    \includegraphics[width=\textwidth]{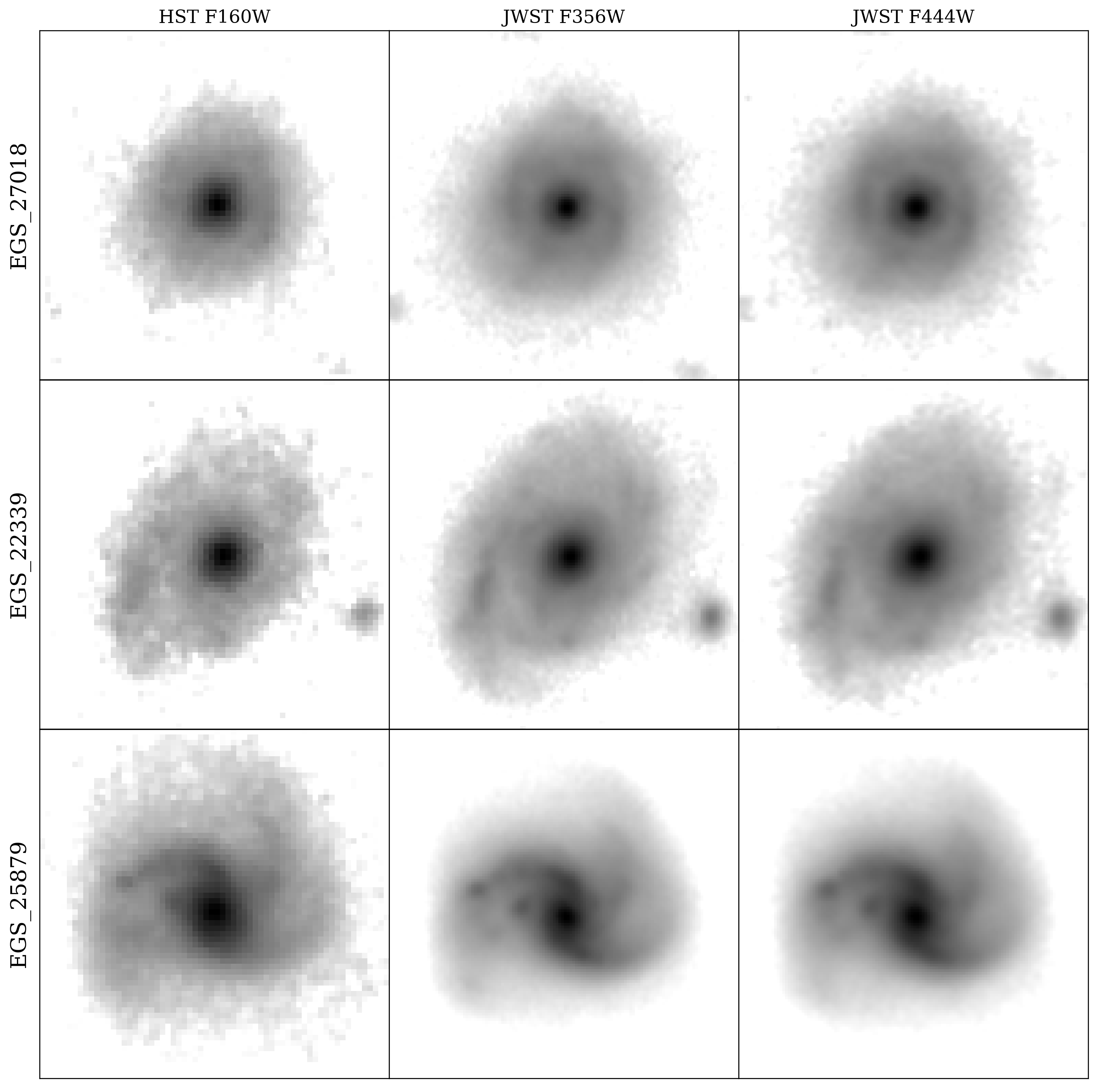}
    \caption{Logarithmic images of the galaxies EGS\_27018, EGS\_22339 and EGS\_25879, which received maybe-barred or barred votes by the classifiers in the HST WFC3 F160W filter, but unbarred votes in the JWST NIRCam F444W filter. The three galaxies are shown in the HST WFC3 F160W (left), JWST NIRCam F356W (middle) and JWST NIRCam F444W (right).}
    \label{Fig: HSTvotes}
\end{figure}

The improved sensitivity and longer wavelength range of JWST enhance galaxy images as compared to what was previously seen in HST WFC3 images. Figure \ref{Fig: HSTvotes} shows three galaxies in the HST WFC3 F160W filter and two JWST NIRCam filters, F356W and F444W. These galaxies are interesting as they received maybe-barred or barred votes in visual classifications of HST WFC3 F160W (for a description of the method, see \S~\ref{Sec: identification}), but unbarred votes from all classifiers in JWST NIRCam F444W. A PSF artefact (as described in \S~\ref{sec:reduction}) can be identified in the JWST NIRCam images of EGS\_27018 but becomes inconspicuous in the HST WFC3 image. EGS\_22339 is a disc galaxy with spiral features, which could have misled visual classifications in the HST WFC3 image but is clearly unbarred in JWST NIRCam images. The only galaxy to receive barred votes in the HST WFC3 filter and unbarred votes in JWST NIRCam filters is EGS\_25879, which is due to the blurring of prominent spiral arms in the F160W filter.

\bsp	
\label{lastpage}
\end{document}